\documentclass[final,3p,times,twocolumn]{elsarticle} 

\usepackage{amssymb}
\usepackage{booktabs}
\usepackage{subcaption}
\usepackage{siunitx}
\usepackage{hyperref}
\usepackage{xcolor}
\usepackage{fixltx2e} 

\usepackage{multicol}
\newcommand\mean[1]{\overline{#1}}

\journal{Nuclear Instruments and Methods in Physics Research Section A}

\begin{document}
\begin{frontmatter}
\title{Commissioning of a dual-phase xenon TPC at Nikhef}

\author{E. Hogenbirk\corref{cor1}}
\ead{ehogenbi@nikhef.nl}
\author{J. Aalbers}
\author{M. Bader}
\author{P.A. Breur}
\author{A. Brown}
\author{M.P. Decowski}
\author{C. Tunnell}
\author{R. Walet}
\author{A.P. Colijn}

\address{Nikhef, Science Park 105, 1098 XG Amsterdam, The Netherlands}

\begin{abstract}
A dual-phase xenon time-projection chamber was built at Nikhef in Amsterdam as a direct dark matter detection R\&D facility.
In this paper, the setup is presented and the first results from a calibration with a $^{22}$Na gamma-ray source are presented.
The results show an average light yield of (\num[separate-uncertainty = true]{5.6 \pm 0.3})~photoelectrons/\si{keV} (calculated to \SI{122}{keV} and zero field) and an electron lifetime of \SI[separate-uncertainty = true]{429 \pm 26}{\micro s}.
The best energy resolution $\sigma_E/E$ is (\num[separate-uncertainty = true]{5.8 \pm 0.2})\% at an energy of \SI{511}{keV}.
This was achieved using a combination of the scintillation and the ionization signals.
A photomultiplier tube gain calibration technique, based on the electroluminescence signals occurring from isolated electrons, is presented and its advantages and limitations are discussed.
\end{abstract}

\begin{keyword}
xenon \sep TPC \sep gain calibration \sep PMT \sep single electron \sep dark matter
\end{keyword}

\end{frontmatter}

\section{Introduction}
There is considerable evidence from astrophysical observations that there is more mass in the universe than can be accounted for with only standard model particles \cite{Clowe:2006eq,planck,structure}.
The most popular theory that explains this discrepancy introduces dark matter particles called WIMPs \cite{Jungman1996195}.
In past years, the sensitivity of direct dark matter search experiments has increased by orders of magnitude, lead by the development of large dual-phase xenon time-projection chambers~(TPCs) \cite{RevModPhys.82.2053, 1748-0221-8-04-R04001, Baudis201450}.
In the context of dark matter research, a small-scale liquid xenon TPC, called XAMS (Xenon Amsterdam), has been designed, built and commissioned at Nikhef in Amsterdam.
The setup described in this work is similar to small-scale dual-phase xenon setups, such as described in \cite{1344272, PhysRevLett.97.081302, Kwong2010328}.

Dual-phase TPCs detect a particle interaction using two distinct signals.
The first comes from excitations and recombined electron-ion pairs.
Bound excited states of two atoms form, and subsequent decays of these excitons causes scintillation light that is detected by photomultiplier tubes (PMTs).
This signal is called S1.
The second signal is caused by ionization electrons that do not recombine with ions.
These are drifted up by an electric field and extracted by a second, stronger field into the gas phase, where secondary scintillation (S2) is caused and measured by the same PMTs.
The drift time between these signals is proportional to the interaction depth (z).
In addition to this, the ratio of S2/S1 provides a powerful discrimination between electronic and nuclear recoils.
In large-scale TPCs, such as XENON100 \cite{Aprile:2011dd} and LUX \cite{Akerib:2012ys}, the light distribution of the S2 in the PMTs gives the coordinates in the plane of the PMTs, so that a three-dimensional resolution is obtained.

This article has the following structure.
In section~\ref{SEC2}, the XAMS setup and the TPC are introduced.
Section~\ref{SEC3} discusses the data processing and gives results based on the main S1- and S2-signals.
In section~\ref{SEC4}, S2-signals from single electrons are analyzed and a PMT calibration technique based on these signals is presented.
In section~\ref{SEC5}, we give a summary of the analyses in these sections.

\begin{figure*}[th]
\begin{center}
\includegraphics[width= 0.95 \linewidth]{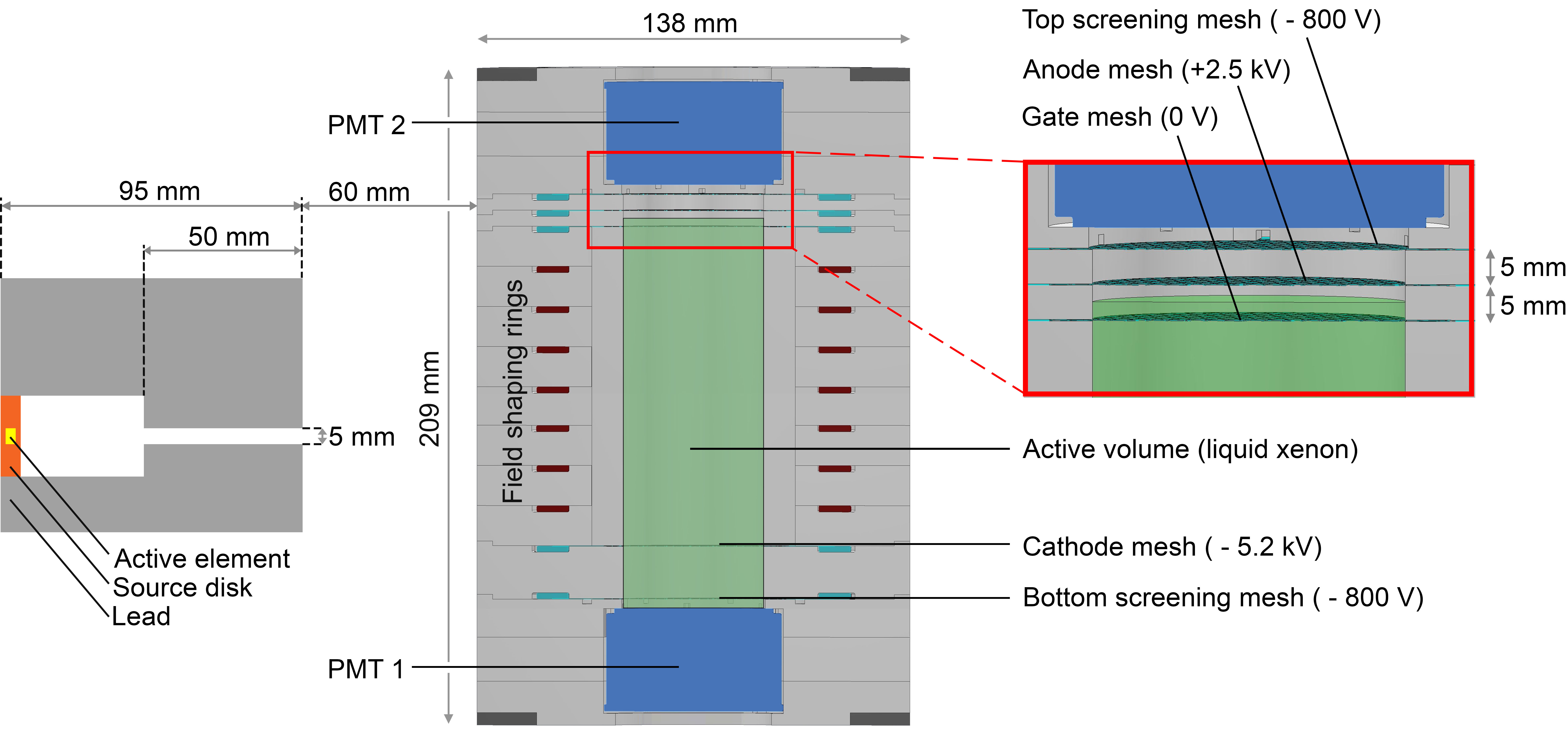}
\caption{
Cross-section of the XAMS TPC and the source in the collimator (drawn to scale).
All elements of the TPC are contained in a cylindrical PTFE structure made from stackable disks, as indicated by the gray color.
The electric field is defined by five meshes and seven copper rings, serving to homogenize the drift field.
The top and bottom screening meshes are held at the cathode potential of the PMTs.
The active volume is defined by the cylindrical volume between the gate and cathode mesh, measuring \SI{100}{mm} (height) $\times$ \SI{44}{mm} (diameter).
The $^{22}$Na gamma source (described in section~\ref{SEC2_3}) is mounted in a collimator that is made of two lead blocks with cylindrical holes.
These are positioned on the outside of the outer vessel (vessels not shown) and allow for a beam size of \SI{11}{mm} at the closest edge of the active volume.
A two-inch NaI(Tl)-detector (not depicted) is used for triggering, and is positioned \SI{100}{mm} to the left edge of the collimator.
The z-position of the collimator is adjustable.
}
\label{FIG1}
\end{center}
\end{figure*}

\section{The XAMS setup}\label{SEC2}
\subsection{The XAMS TPC}
The XAMS TPC features a cylindrically-shaped active volume of \SI{154}{cm^{3}}, which holds \SI{434}{\gram} of liquid xenon at a temperature of \SI{-90}{\celsius}, as shown in Fig.~\ref{FIG1} \cite{erik_master,maria_master,rolf}.
There are two PMTs, one at the top and one at the bottom, that view the active volume and record the S1- and the S2-signals.
Five meshes define the electric field: the drift field of \SI{0.52}{kV/cm} is between the cathode and the grounded gate mesh, whereas the extraction field  is between the gate and anode mesh, where a voltage of \SI{2.5}{kV} is applied over \SI{5}{mm}.
The meshes were made by chemical etching of a \SI{150}{\micro m} thick stainless steel sheet.
They have a square pattern with a pitch of \SI{2.45}{mm} and a wire thickness of \SI{150}{\micro m}, giving a head-on optical transparency of \num{88}\%.
The drift field is shaped by a series of copper rings connected to a resistor chain between the cathode and gate mesh.
Two additional meshes shield the PMTs from the TPC's electric fields.
The distance between the cathode and the gate mesh, which defines the maximum drift length, is \SI{100}{mm}.

The PMTs are circular two-inch UV-sensitive low-temperature Hamamatsu PMTs of type R6041-406.
The low transit-time spread of \SI{0.75}{ns} in combination with the fast 500~MSa/s digitizer type CAEN V1730D makes XAMS well-suited for fast-timing applications, such as pulse-shape discrimination studies.

\subsection{Cryogenics and gas system}
For the successful operation of a dual-phase xenon TPC, a cryogenic cooling system is required in combination with a purification and storage system.
The piping and instrumentation diagram of the XAMS setup is included in~\ref{APP_A}.

The cryogenic part of the system consists of double-walled stainless steel vessels.
The insulation volume between the vessels is continuously pumped out during normal operation, and pressures of \SI{3e-7}{mbar} are reached. 
In addition, aluminum-coated Mylar foil is inserted in the insulation volume to shield from radiative heat transfer.
The cooling is provided by an Iwatani PDC08 pulse tube refrigerator (PTR), which gives an effective cooling power of \SI[separate-uncertainty = true]{22 \pm 2}{\watt} at \SI{-90}{\celsius}.
We apply the cooling to a copper cold finger, where the xenon condenses and droplets fall down into a funnel leading into the TPC.
A resistive heating band wrapped around the cold finger enables us to regulate the temperature.
The current to the heating band is controlled by a PID controller based on the temperature read by a Pt100 temperature sensor at the cold finger.

A cooling power failure may result in a rising pressure in the TPC.
A burst valve with a pressure limit of $\sim$~\SI{4.0}{bar} is connected to the inner volume to ensure no higher pressure can build.
We provide emergency cooling with a pressurized liquid nitrogen dewar, with the flow controlled by a solenoid valve that is switched by a pressure sensor.
In addition, text and email warning messages are automatically sent in case of abnormal behavior of the system.
The pressure sensor, the solenoid valve and the computer that sends the messages are powered by an uninterruptible power supply.

The required xenon purity level is achieved by continuous circulation through a high-temperature SAES MonoTorr PS3-MT3-R-2 getter with a maximum flow rate of 5~standard liters per minute.
We use a heat exchanger at the cryogenic part of the system to achieve this flow rate with only modest cooling power.
In section~\ref{SEC3_2}, we show that we achieved an impurity level of  \SI[separate-uncertainty = true]{1.2 \pm 0.1}{ppb} (oxygen-equivalent).

We use an EMP MX-808ST-S diaphragm pump to establish the flow in the recirculation circuit.
The flow is controlled with a needle valve and measured with a thermal mass flow meter.
For the measurements described in this work, no buffer volumes were installed at the inlet or outlet of the pump, causing oscillatory behavior in the flow.
The presumed effect on the measurements is described in detail in section~\ref{SEC3_3_1}.
We recognize this as a design flaw, which we have since adjusted by installing gas bottles as buffer volumes in the system.

The liquid level in the TPC is monitored by a stainless steel cylindrical capacitive level meter, which is read out by a custom-programmed Arduino board.
The flow control of the needle valve is used to set the liquid level, as we noticed that the liquid level decreased as we increased the flow rate.
We assume that this effect is due to a changing thermal equilibrium in the heat exchanger, where a nonnegligible amount of liquid xenon is kept.

The total xenon content in the XAMS setup is roughly \SI{6}{kg}, most of which surrounds the PTFE structure of the TPC.
The time required to fill the TPC, limited by the maximum cooling power of the PTR, is roughly \SI{10}{hours}.
We perform recuperation by immersing gas bottles into liquid nitrogen dewars and allowing gas to deposit on the walls of the cylinder.
The time for a full recuperation is roughly \SI{8}{hours}.

\subsection{Trigger and DAQ} \label{SEC2_3}
We use a $^{22}$Na gamma source with an activity of \SI[separate-uncertainty = true]{368 \pm 11}{kBq} to perform our studies.
The source is mounted in a lead collimator (see Fig.~\ref{FIG1}) on the outside of the insulation vacuum vessel, with an opening angle of \SI{2.9}{\degree} such that the beam has a width of \SI{11}{mm} at the closest edge of the active volume.
The direction of the beam is horizontal, giving lateral irradiation of the TPC.
We change the z-position of the collimator by varying the height of the platform on which the collimator is mounted.
To reach the active volume, the gamma rays have to cross the walls of the inner and outer vessels, a thin layer of liquid xenon and the PTFE holding structure of the TPC, so that the total material traversed is \SI{6}{mm} of stainless steel, \SI{2}{mm} of liquid xenon and \SI{46}{mm} of PTFE, respectively.

$^{22}$Na decays by positron emission (branching ratio 90.4\%) or electron capture (branching ratio 9.6\%). 
The decay is almost always followed by the emission of a \SI{1274}{keV} gamma ray from its $^{22}$Ne daughter.
In the case of positron emission, two additional back-to-back gamma rays of \SI{511}{keV} are produced from positron annihilation.
By using thallium-doped sodium iodide~(NaI(Tl)) as a coincidence detector that measures one of the \SI{511}{keV} gamma rays, the other \SI{511}{keV} gamma ray going directly toward the active volume is tagged.
This increases the fraction of events where all the energy is absorbed, since the number of events where gamma rays enter the active volume after Compton scattering on the material surrounding the detector is reduced.

The trigger is based on a threefold coincidence of the two PMTs in the TPC and the external NaI(Tl) detector.
If the trigger condition is satisfied, all three channels are digitized by a CAEN V1730D digitizer board.
This board has 8 channels that are digitized with a time resolution of \SI{2}{\nano s} and a voltage resolution of \num{14} bits, distributed over a dynamic range of \SI{2}{V}.
We choose an event window of \SI{163}{\micro s}: more than twice  as long as the maximum drift time of \SI{60}{\micro s}.
We place the trigger position in the middle, such that an (accidental) trigger on an S2-signal will always contain the S1 in the same window.
A cut in post-processing ensures that there was a true coincidence with the S1 and the external NaI(Tl) (and not, for example, a coincidence with the S2-signal and an uncorrelated interaction in the NaI(Tl) crystal).

The simple coincidence means that all three channels must exceed the threshold \emph{at the same time}; no coincidence window was used.
The time offset between the two PMTs in the TPC is negligible, however, the start of the peak of the NaI(Tl) detector output was shown to occur \SI[separate-uncertainty = true]{22 \pm 6}{ns} later than that of the PMT signals.
The trigger condition was therefore satisfied only if both PMT signals were still above threshold at this time after the peak amplitude.
In the case of high energy recoils, the pulses are sufficiently large and this causes no problems.
However, for low energy recoils we observe a low trigger efficiency, which we identify in the comparison to Monte Carlo simulation in section~\ref{SEC3_3_2} at energies below~\SI{150}{keV}.

\section{Data reduction and results} \label{SEC3}
\subsection{Peak finding, clustering, identification} \label{SEC3_1}
\begin{figure}[h]
\begin{center}
\includegraphics[width=\linewidth]{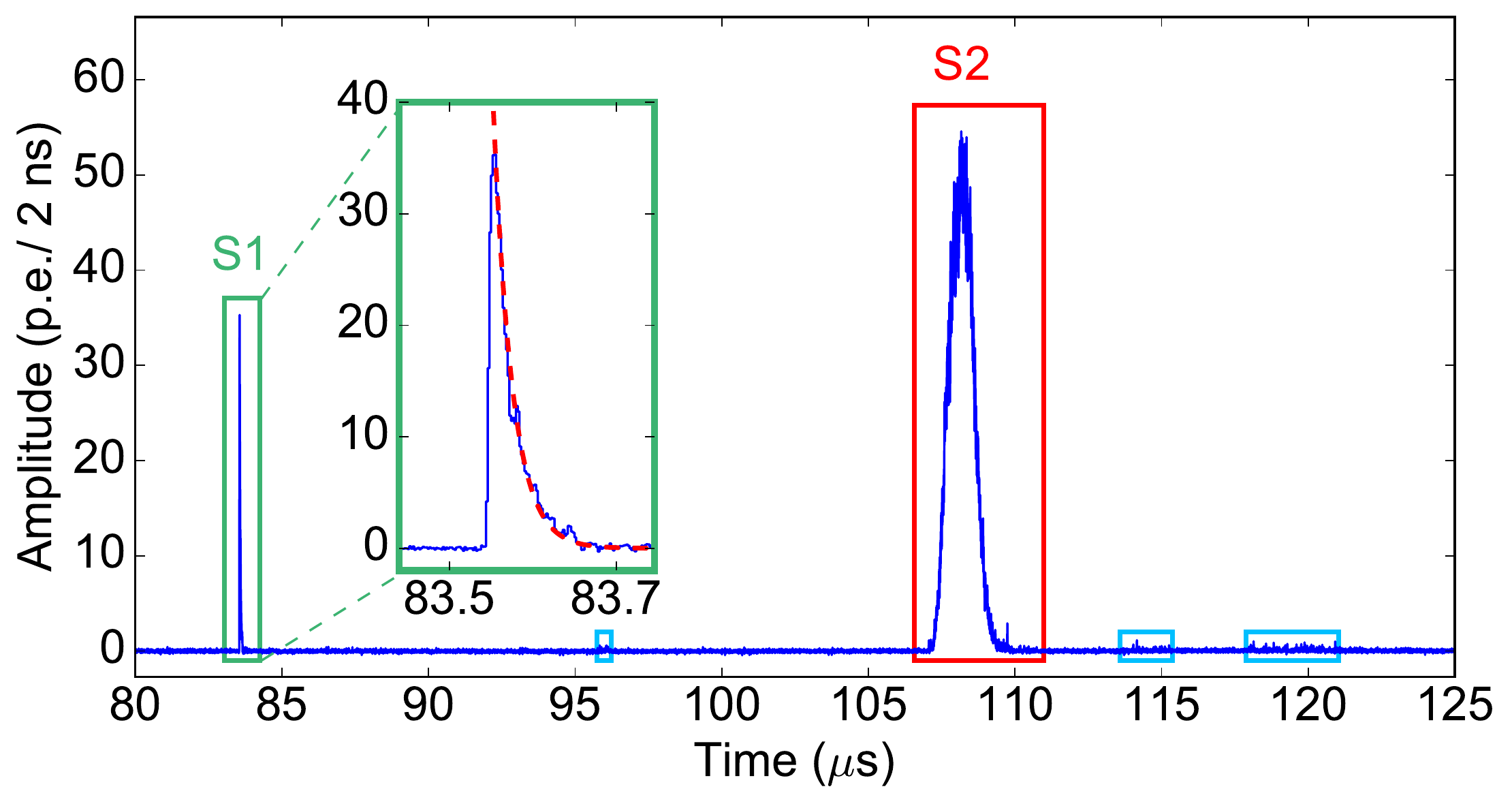}
\caption{Typical gamma-ray-induced sum-signal of the two PMT channels, showing the S1 (green box) due to prompt scintillation and the S2 (red box), delayed by the drift time. 
The inset shows a detailed view of the S1-signal and an exponential fit to the falling slope of the S1, with a decay time of \SI[separate-uncertainty = true]{22.8 \pm 0.1}{\nano s}.
Additional peaks are found (in the blue boxes), mostly happening after the S2.
Details of this kind of signal can be found in Fig.~\ref{FIG7}.
The data processing software finds the hits in each channel, clusters them into peaks, determines peak properties and classifies each peak based on the width and area.
}
\label{FIG2}
\end{center}
\end{figure}

The data of each event consist of the waveforms of the two PMTs with a duration of \SI{163}{\micro s} (Fig.~\ref{FIG2}).
The data processor, which is the same software developed for XENON1T \cite{pax_url}, analyzes the waveforms in each individual PMT channel by looking for significant excursions above the baseline.
These are called \emph{hits}.
In XENON100 and XENON1T, \emph{zero-length encoding} is used: the data consists of small chunks of data around a significant excursion from the baseline, so that the baseline is suppressed and the data volume is reduced \cite{zle}.
In order to be compatible with this structure, we apply a software zero-length encoding with a very low threshold.
The hitfinder threshold is dynamically determined as \num{4.5} times the standard deviation of the noise in the first 40 digitizer samples (\SI{80}{ns}) of the zero-length encoded chunk containing the hits.
The hits from both channels are then clustered into \emph{peaks} based on the gap between the edges of the hits: if this exceeds \SI{450}{\nano s}, the hits are clustered into separate peaks.
The area and the width of the peak are computed based on the summed gain-corrected waveform properties.
The width metric uses the range containing \num{50}\% of the peak area with \num{25}\% on either side. 
The peak position is defined as the amplitude-weighted mean time of the samples in the peak.

As seen in Fig.~\ref{FIG2}, the main signals are the S1- (highly peaked signal at~\SI{84}{\micro s}) and the S2-signal (the broad signal at~\SI{108}{\micro s}).
After the S2, some peaks with low area and high width are found (shown in blue boxes).
These signals are due to secondary emission of electrons caused by photo-ionization of S2 UV photons and drifted up to produce an additional, much smaller S2. 
These signals will be discussed in section~\ref{SEC4}.
All peaks are classified as either `S1', `S2' or `other' based on their width and area.

\subsection{S1 and S2 corrections} \label{SEC3_2}

After data processing, the following selection criteria are applied to the events.
First of all, only events with a single S1 and S2 are kept.
This cut rejects pileup events, double scatter events (which cause two S2-signals), or events where no S2 is generated (for instance, where the interaction occurs below the cathode mesh).
Events where the S1 is not in coincidence (difference of peak center position less than \SI{200}{ns}) with a signal in the NaI(Tl)-crystal are also cut.
In addition, the energy deposition in the NaI(Tl)-crystal is required to be less than \SI{600}{keV}, so that the triggers on \SI{1274}{keV} gamma ray are cut.
We impose no lower bound other than the trigger threshold on the NaI(Tl) energy, so that we keep events where the \SI{511}{keV} Compton scattered in the NaI(Tl)-crystal.
A summary of all the cuts and the number of events surviving each successive cut is given in table~\ref{TABLE1}.
The fraction of events surviving all cuts for the analysis presented here is 47.0\%.
Most events cut are due to multiple S1s or S2s.

\begin{table}[tp]
\caption{Data selection cuts and the number and fraction of events surviving each cut.
The cuts are applied successively.}
\label{TABLE1}\centering
\begin{tabular}{lcc}
\toprule%
{\bf Cut }  & {\bf Events} & {\bf Fraction}   \\ \toprule
No cuts & \num{215831} & 100.0\% \\ 
At least one S1 & \num{205417} & 95.2\% \\
Only one S1 & \num{166353} & 77.1\% \\
At least one S2 & \num{158586} & 73.5\%\\
Only one S2 & \num{115005} & 53.3\%\\
Coincidence S1 and NaI(Tl) & \num{105062} & 48.7\%\\
NaI(Tl) $<$ \SI{600}{keV} & \num{101381} & 47.0\%\\
\midrule 
{\bf Total} & {\bf 101~$\!$381} & {\bf 47.0\%} \\
 \bottomrule
\end{tabular}
\end{table}

\begin{figure*}[t]
\centering
\begin{subfigure}[b]{.5\linewidth}
\centering \includegraphics[width= 1.0 \linewidth]{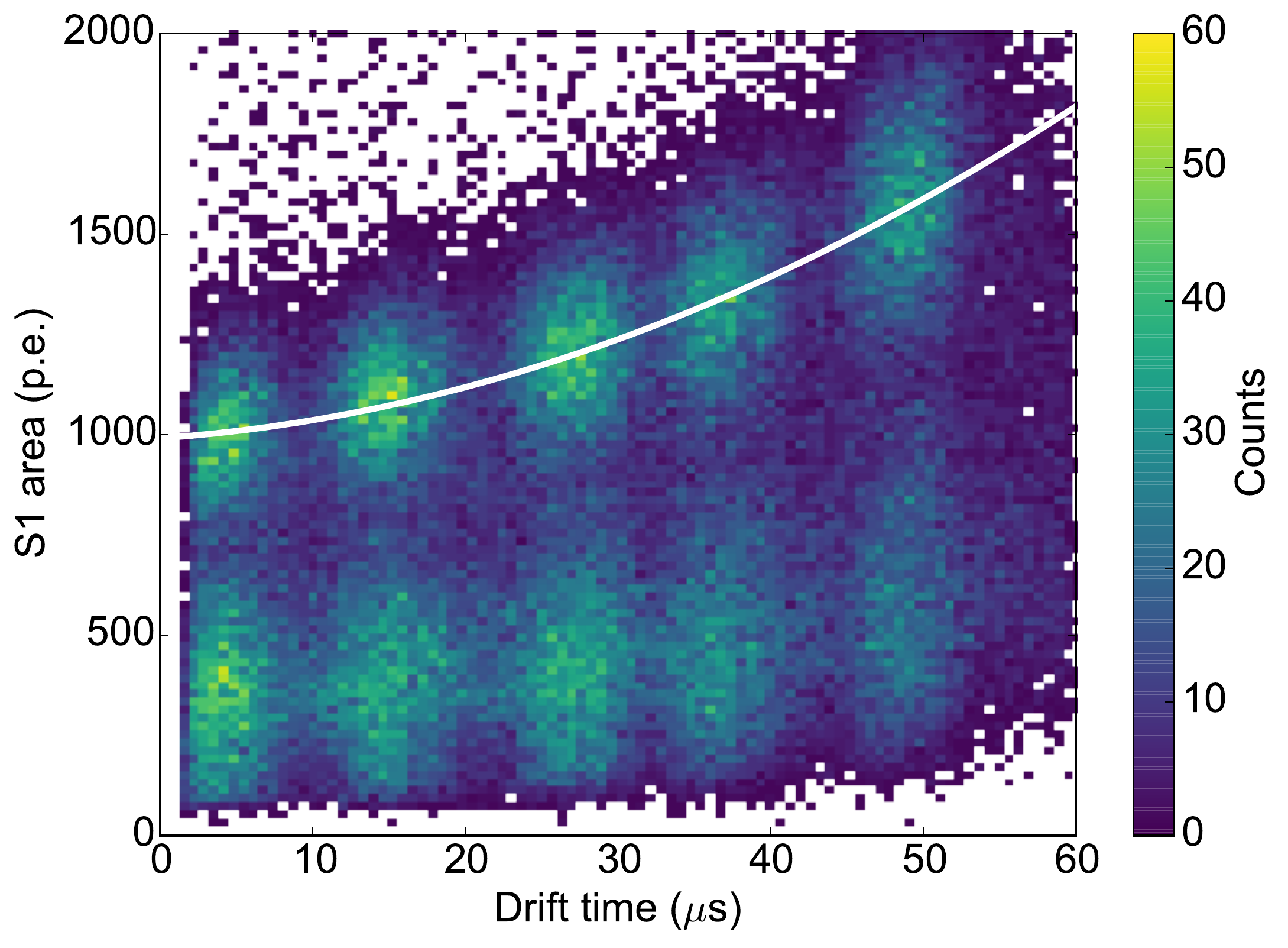}
\caption{ }\label{FIG3A}
\end{subfigure}%
\begin{subfigure}[b]{.5\linewidth}
\centering \includegraphics[width=1.0 \linewidth]{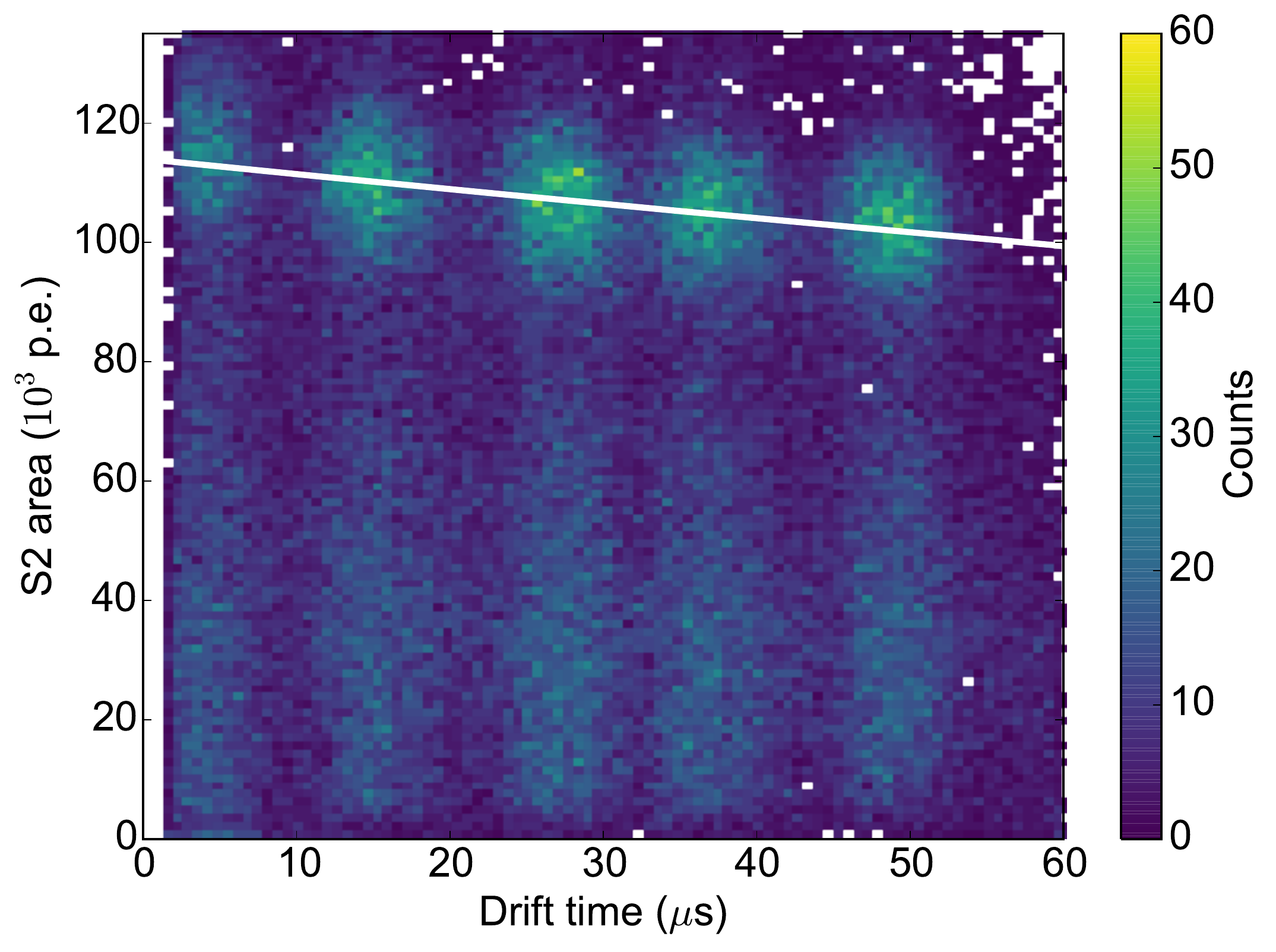}
\caption{ }\label{FIG3B}
\end{subfigure}
\caption{Density plot of the area of the sum-signal of the S1 {\bf (a)} and the S2 {\bf (b)} signal for different z-positions in the TPC, corresponding to different drift times.
The data shown in these figures were taken with the collimated source pointing at five different positions in the TPC.
The thick white lines are fits to the photo-peak.
For the S1, an overall increase is found due to LDE effects, a second degree polynomial fit gives the correction.
For the S2, an exponential fit provides a correction for loss of electrons during the drift time.
}
\label{FIG3}
\end{figure*}

For both the sum-signals of the S1- and the S2-signals, the area of the peak is proportional to the recoil energy.
However, the response to a mono-energetic energy deposition is not uniform throughout the TPC, requiring spatial corrections.
Since the XAMS TPC has only two PMTs, the position in the x,y-plane cannot be determined, but the z-coordinate is calculated based on the drift time that is defined by the difference of the weighted mean times of the S2 and the S1.

\begin{figure*}[t]
\centering
\begin{subfigure}[b]{.5\linewidth}
\centering \includegraphics[width= \linewidth]{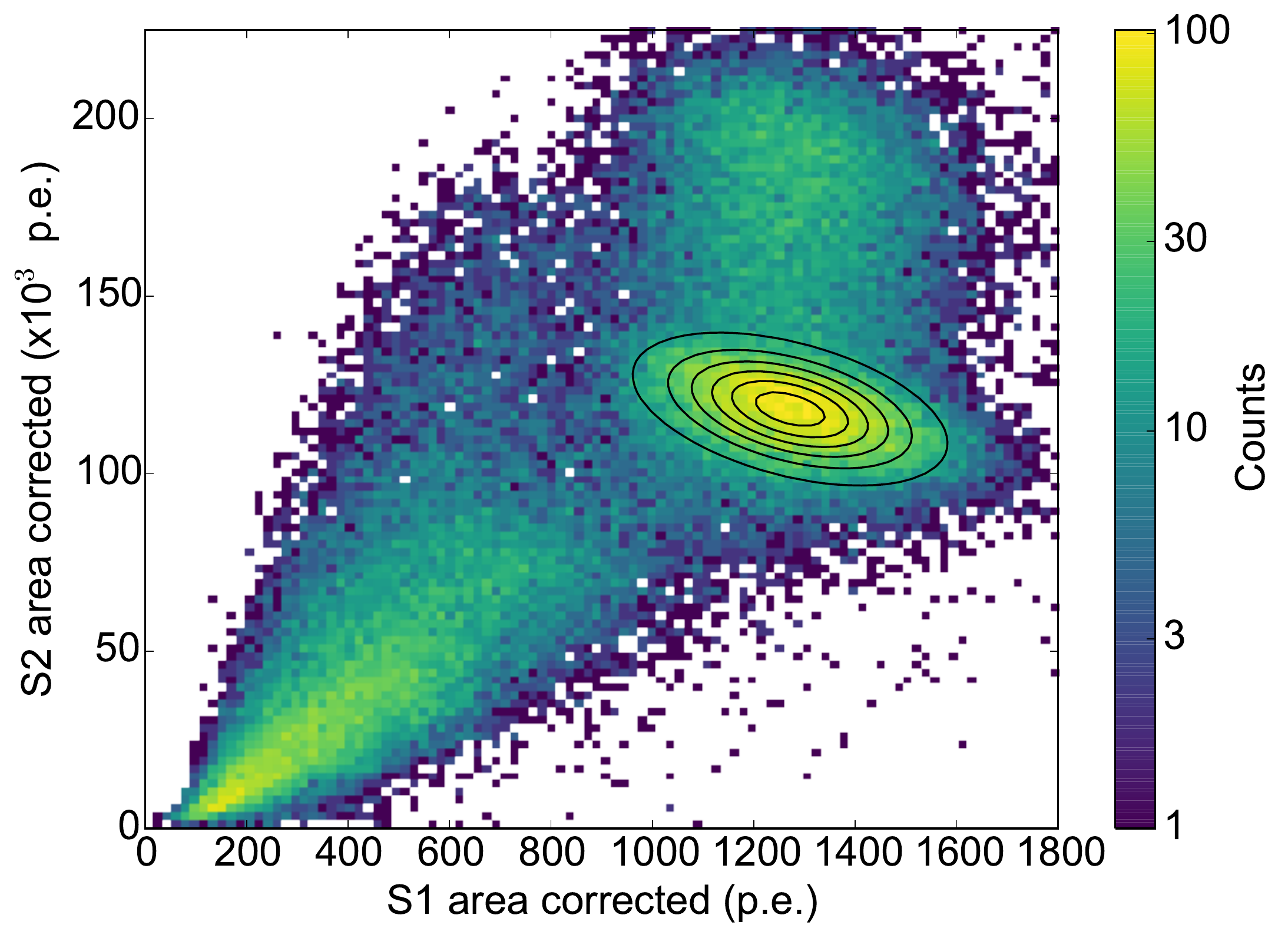}
\caption{ }\label{FIG4A}
\end{subfigure}%
\begin{subfigure}[b]{.5\linewidth}
\centering \includegraphics[width= \linewidth]{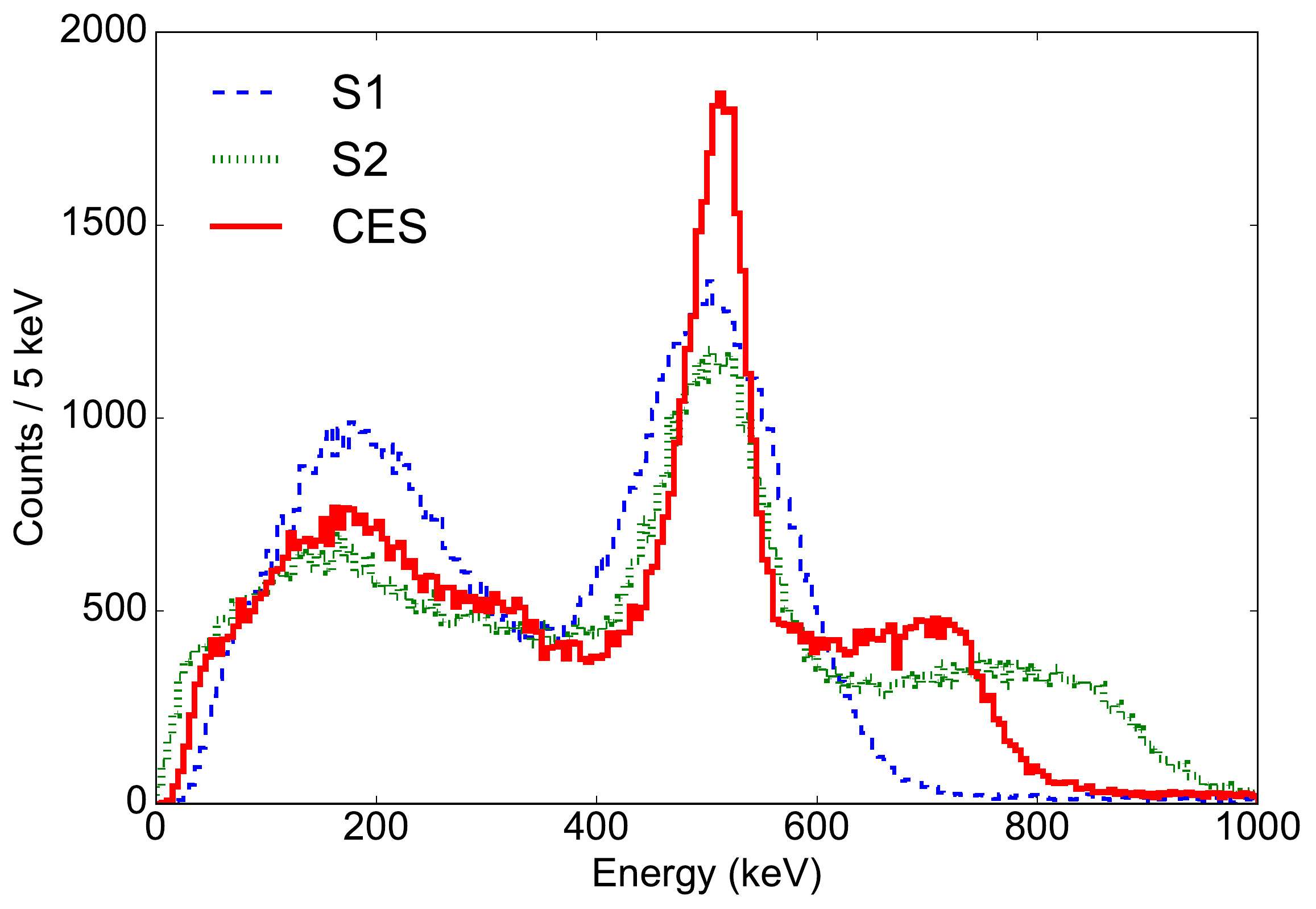}
\caption{ }\label{FIG4B}
\end{subfigure}
\caption{{\bf (a)}: Density plot of the area of the S1 and the area of the S2 in the same event for a \SI{511}{keV} gamma-ray source. 
The ellipse shows the anti-correlation between the S1- and the S2-signal at the expected photo-peak.
A superior energy resolution is found by fitting the photo-peak and projecting along the short axis of the ellipse.
The shoulder at low energy is due to Compton-scatter events; the second peak at higher S2 area than the photopeak is discussed in the text.
{\bf (b)}: The spectra using the S1, S2 and the combined signal.
The energy resolution at \SI{511}{keV} improves from (\num[separate-uncertainty = true]{14.5\pm 0.2})\% and (\num[separate-uncertainty = true]{10.8\pm 0.4})\%, respectively, for the S1- and the S2-signal alone to (\num[separate-uncertainty = true]{5.8\pm 0.2})\% for the combined spectrum.
}\label{FIG4}
\end{figure*}

A z-dependent scale factor is applied to the S1-signal to eliminate differences in light detection efficiency (LDE). 
The amount of light detected by the PMTs for different interaction positions depends on optical properties of the TPC, such as the reflection properties of the walls of the TPC, optical transparency of the meshes and reflection on the liquid-to-gas interface.
The secondary scintillation light of the S2-signal is always produced in the small region between the liquid-to-gas interface, so no z-correction for LDE  has to be applied.
However, the number of electrons that create the S2-signal decreases with increasing drift time due to attachment of electrons to impurities in the xenon.
Assuming~$n_0$ electrons are produced at the interaction position, the number of electrons~$n_e$ left after a drift time~$t_d$ can be calculated with
\begin{equation}
n_e = n_0 \exp{\left(-t_d/ \tau_e\right)} {\rm ,}
\label{EQN1}
\end{equation}
where $\tau_e$ is the \emph{electron lifetime}, which is an indirect measure of the purity of the xenon.

Five datasets were taken with a different z-position of the collimator.
Fig.~\ref{FIG3} shows the area of the sum-signal of the S1- and the S2-signal for all datasets, each containing  a prominent peak at high energy and a broad shoulder for lower energies.
The former is attributed to the full absorption peak (mostly due to photoelectric absorption, or multiple scatter events where the S2s are too close together to be separated), whereas the latter is due to Compton-scatter events.

For the S1, uncertainties on optical parameters limit the use of a detailed LDE model.
We therefore use a data-driven approach, modeling the correction function as a second degree polynomial.
We determine this function in two steps.
We first fit a Gaussian function to the photopeak for several slices in drift time, and then fit the photopeak position as a function of drift time with a second-degree polynomial.
This polynomial function, shown by the white line in Fig.~\ref{FIG3A}, provides the correction factor for the LDE.
The average value of the fit function, which gives the volume-averaged light yield for \SI{511}{keV} gamma rays, is \num[separate-uncertainty = true]{1.29 \pm 0.07 e3}~p.e., or (\num[separate-uncertainty = true]{2.5 \pm 0.1})~p.e./keV in this configuration.
This is equivalent to (\num[separate-uncertainty = true]{5.6 \pm 0.3})~p.e./keV at zero field and \SI{122}{keV} using data from NEST \cite{Szydagis:2011tk}, which is comparable to TPCs like XENON100 (4.3~p.e./keV) and LUX (8.8~p.e./keV) \cite{Aprile:2011dd,PhysRevLett.112.091303}.
An overall increase of LDE with drift time is found, since most of the scintillation light is detected by the bottom PMT.

For the S2, the correction function is expected to be an exponential (see equation~\ref{EQN1}).
The electron lifetime as determined from the fit is \SI[separate-uncertainty = true]{429 \pm 26}{\micro s}, similar to the average lifetime of \SI{514}{\micro s} during the year-long science run of XENON100 \cite{xe100_225}, and was achieved in only \SI{7}{days} of continuous purification with the high-temperature getter.
We observed that the electron lifetime rapidly increases in the first \SI{6}{days}, but levels off after this \cite{maria_master}.
For XAMS, this electron lifetime means that even at the maximum drift time, only  13\% of the S2-signal is lost.
Using the values in \cite{noble_gas_detectors}, this electron lifetime corresponds to an impurity level of \SI[separate-uncertainty = true]{1.2 \pm 0.1}{ppb} (oxygen-equivalent).
We kept the recirculation flow rate constant over the full duration of all measurements described here (one day).

\subsection{Energy calibration}
After the corrections for the S1- and the S2-signal have been applied, the absorbed energy can be determined.
Both signals provide a measurement of the deposited energy, since the area of the S1-signal is proportional to the number of photons produced in the interaction and the area of the S2-signal is proportional to the number of electrons produced.
The total energy deposited in these events is always identical: the ionization and scintillation signals are therefore anti-correlated.
In Fig.~\ref{FIG4A}, this anti-correlation is clearly visible as the ellipse with a downward slope.
The best energy resolution is achieved by using a projection along the short axis of the ellipse, which is known as the combined energy scale (CES) \cite{Shutt2007451, PhysRevB.76.014115}.
We use the same projection for all energies, which is a good approximation for energies greater than roughly \SI{100}{keV} \cite{Aprile:2011dd}.
In Fig.~\ref{FIG4B}, the spectra obtained from the S1, the S2 and the CES are shown.
The energy resolutions, as defined by $\sigma_E/E$ for a Gaussian fit, are (\num[separate-uncertainty = true]{14.5\pm 0.2})\%, (\num[separate-uncertainty = true]{10.8\pm 0.4})\% and (\num[separate-uncertainty = true]{5.8\pm 0.2})\%, respectively.

\subsubsection{High-S2 population} \label{SEC3_3_1}
In addition to the photopeak, a second peak at the same S1-area but larger S2-area was found, see Fig.~\ref{FIG4A}.
We also find this effect for the Compton-scatter events, and throughout all datasets.
The appearance of the high-S2 events is highly correlated in time, with a frequency of \SI[separate-uncertainty = true]{0.110 \pm 0.006}{Hz}, i.e., roughly a \SI{9}{s} period (see Fig.~\ref{FIG6A}).

The cause of a varying S2 size can be related to only few parameters.
Since the S1-signal is unaffected, the PMT gain, cathode voltage, DAQ problems or processing errors can be excluded.
Possible detector parameters changing the S2 size, but not the S1 size, are the xenon purity, the anode voltage, and the liquid level.
The anode voltage was not monitored by the slow control system, but the display showed a stability of better than \SI{1}{V}.
Unfortunately, we cannot correlate the detector parameters monitored by the slow control to the time behavior found in the high-S2 population appearance.
This is because the variables were read out only every two minutes; a decision that was taken because the readout of temperature sensors in the TPC caused noise in the PMT signals by electronic pick-up.

\begin{figure}
\begin{center}
\includegraphics[width= \linewidth]{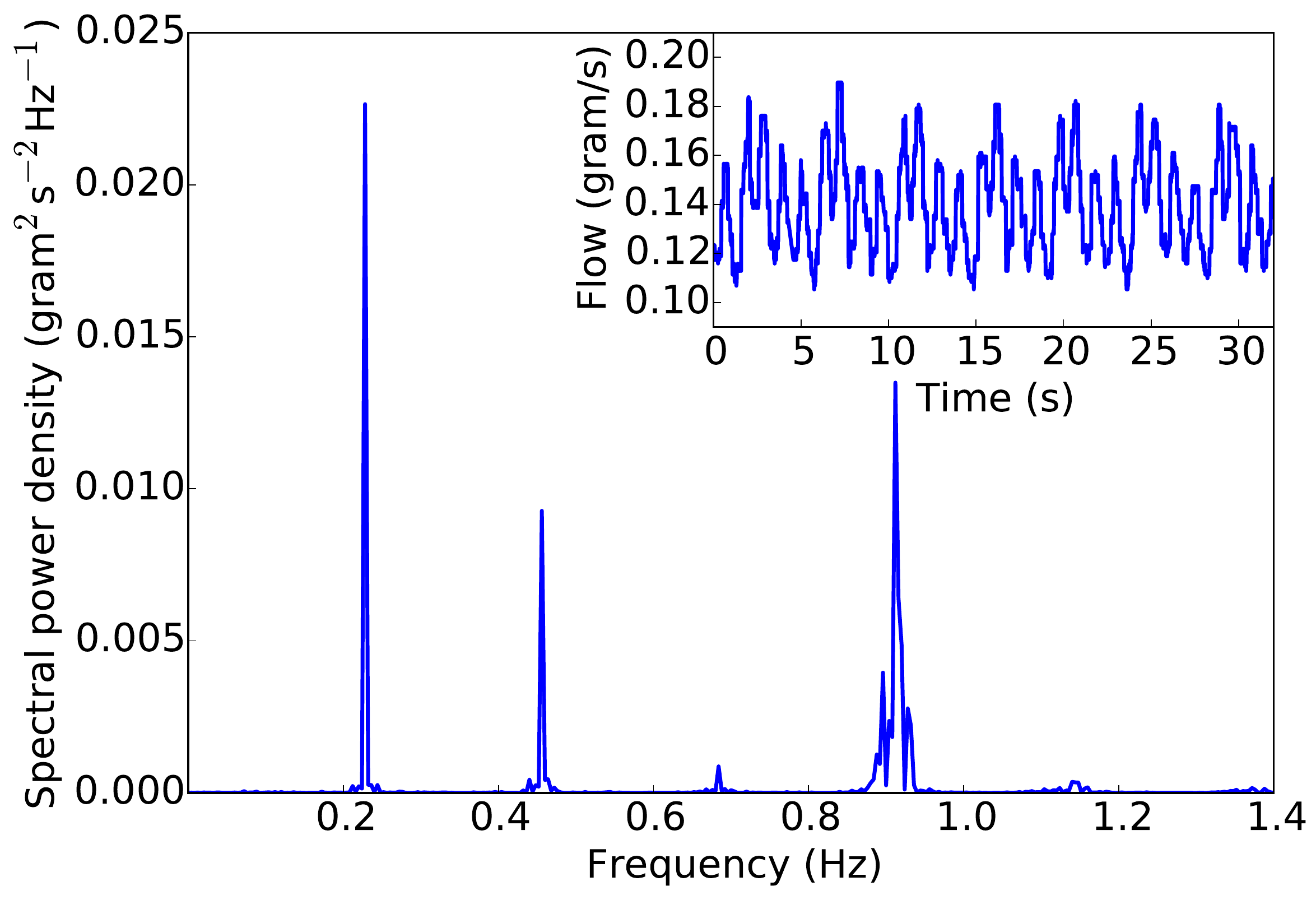}
\caption{Power spectrum of the flow rate as measured in a test where nitrogen gas was pumped through the detector volume.
Shown in the inset is the flow as a function of time for the first \SI{30}{seconds} of this measurement. 
A clear peak at a frequency of \SI[separate-uncertainty = true]{0.228 \pm 0.004}{Hz} is visible, along with several harmonics of this frequency.}
\label{FIG5}
\end{center}
\end{figure}

A plausible explanation found is a time-varying liquid level in the TPC.
The S2 size is highly dependent on this, so that only a small change in liquid level can still give significant effects.
Alternatively, there could be ripples on the liquid surface, appearing every \SI{9}{s}.
One of the mechanisms that could cause either a changed liquid level or ripples on the surface is related to the recirculation flow.
During the measurements, we observed that the gas flow rate in the recirculation system was constantly varying.
To investigate this effect further, we did a test where nitrogen gas was pumped through the system.
We observed a highly periodic behavior of the flow rate, with a period depending on recirculation speed.
Fig.~\ref{FIG5} shows the flow rate for a mass flow similar to the flow used during the measurements with liquid xenon.
The typical frequencies found in these tests are higher than the \SI[separate-uncertainty = true]{0.110 \pm 0.006}{Hz} found in the data, but it should be noted that the systems with liquid xenon and with nitrogen gas are not equivalent, and that the frequency found in the nitrogen gas tests depends on the pressure and the recirculation flow.
The reason for this oscillatory behavior is related to the absence of a buffer volume at the recirculation pump.
Buffer volumes were installed, and subsequent tests showed a significant increase in the stability of the flow rate. 
Future measurements with liquid xenon will show if the effect is related to the instability of the flow rate.

As illustrated in Fig.~\ref{FIG6A}, we can use a time cut to remove a large fraction of the events with a large S2-signal.
Whenever more than six events with an S2 size larger than \num{150000}~p.e.\ are found within one second, the events from one second before to one second after this bin are cut.
This removes \num{41.1}\% of all events passing previous cuts.

\subsubsection{Comparison to simulation} \label{SEC3_3_2}
\begin{figure*}[t]
\centering
\begin{subfigure}[b]{.5\linewidth}
\centering \includegraphics[width= \linewidth]{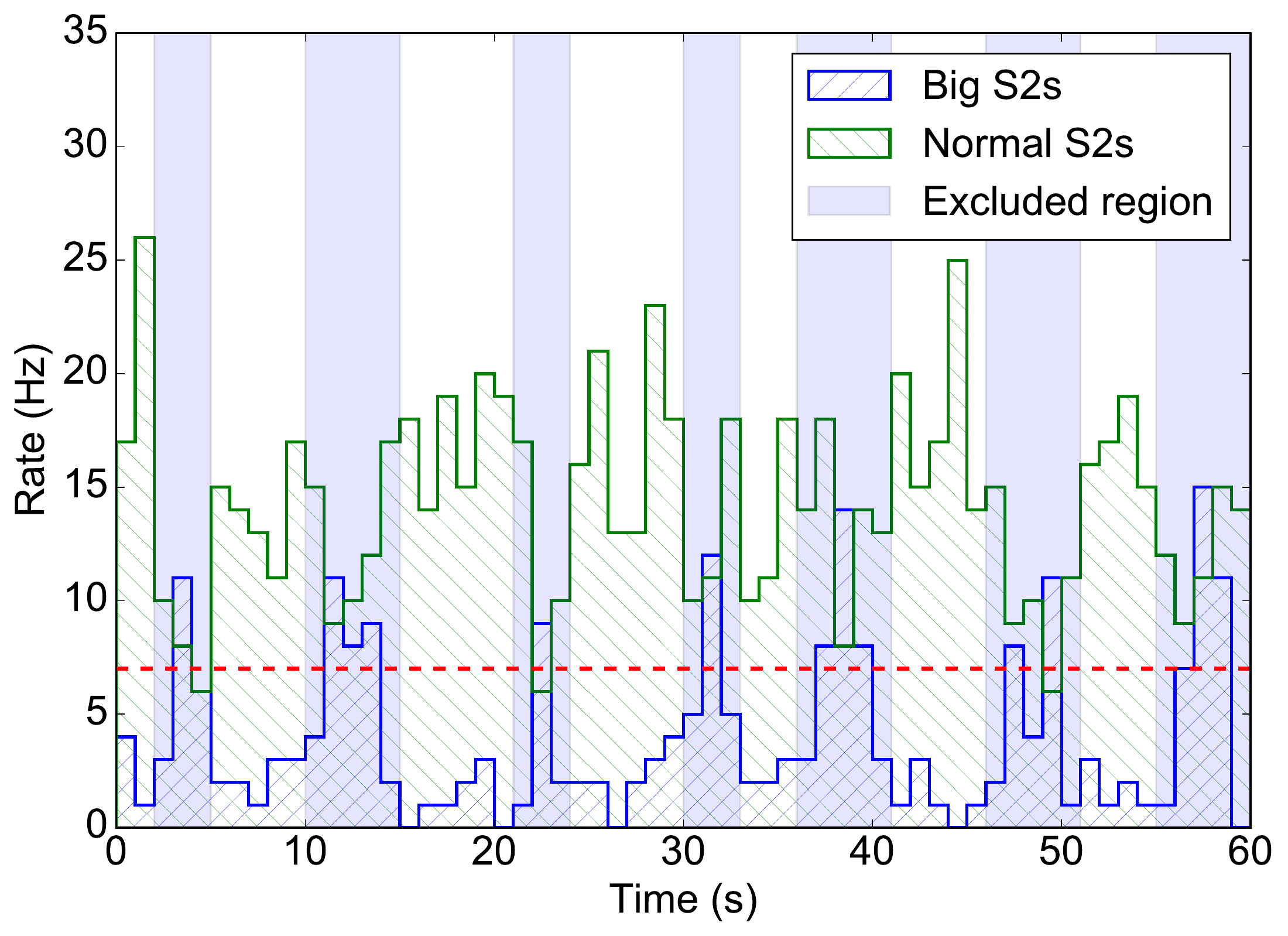}
\caption{ }\label{FIG6A}
\end{subfigure}%
\begin{subfigure}[b]{.5\linewidth}
\centering \includegraphics[width= \linewidth]{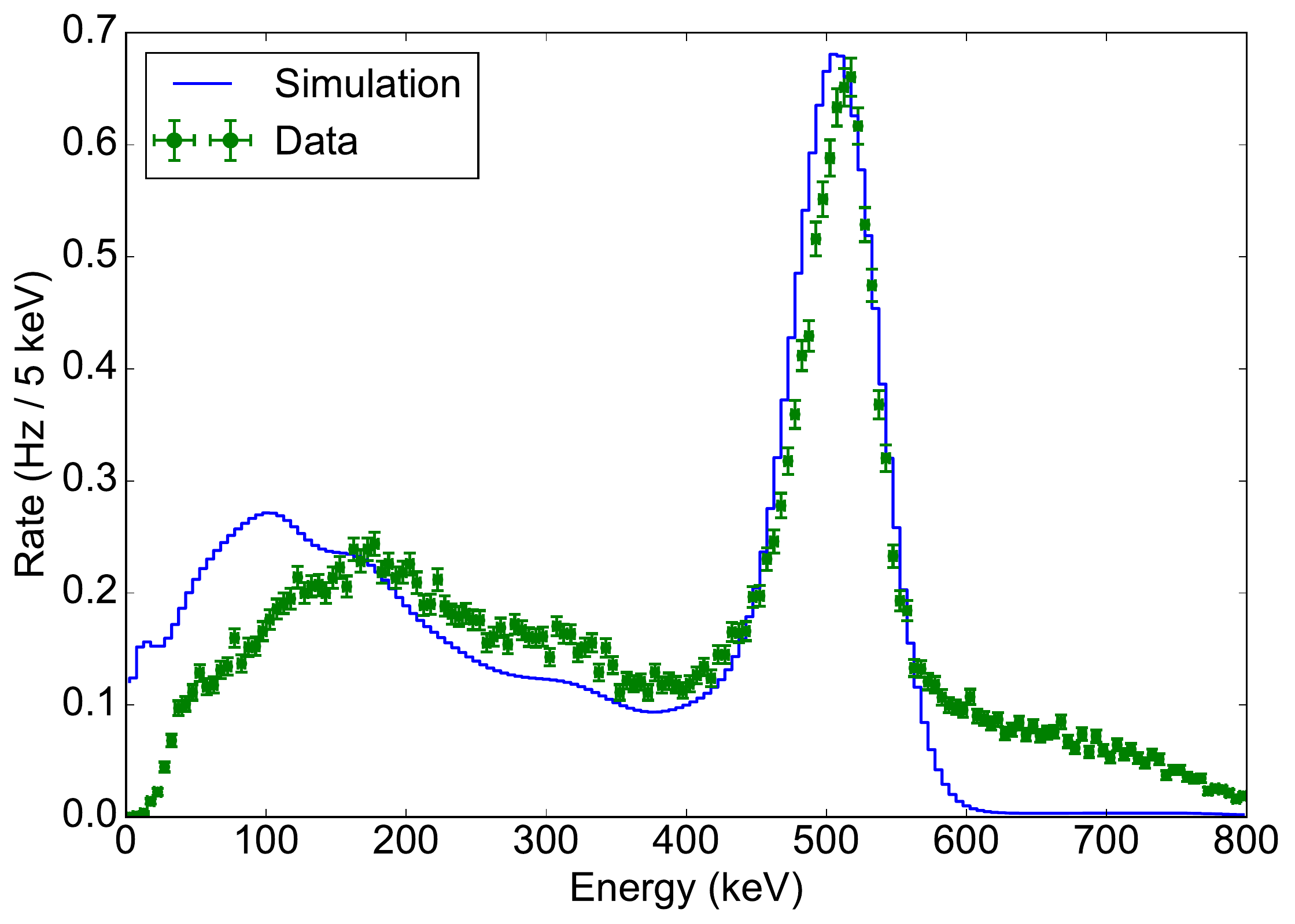}
\caption{ }\label{FIG6B}
\end{subfigure}
\caption{{\bf (a)}: The rate of events with an S2 larger than \num{150000}~p.e.\ (blue) and the rate of other events (green) for the first 60 seconds of the dataset.
A clear time-correlation is visible.
The shaded regions show the times that are cut.
{\bf (b)}: The CES spectrum from data after applying the time cut (green points) compared to a smeared spectrum from a Monte Carlo simulation (blue line). 
The data points are scaled to reflect the rate before any cuts.
The mismatch between simulation and data at low energy is due to a decreased trigger efficiency, as described in section~\ref{SEC2_3}.
At high energy, this is due to the partly cut high-S2 population described in section~\ref{SEC3_3_1}.
}\label{FIG6}
\end{figure*}

The resulting spectrum was compared to a GEANT4 \cite{geant4} Monte Carlo simulation, where the energy deposition was registered when there is a simultaneous energy deposition in the NaI(Tl) crystal and the liquid xenon active volume.
The result was then smeared with an energy resolution function according to

\begin{equation}
\frac{\sigma_E}{E} = \frac{a}{\sqrt{E}} {\rm ,}
\end{equation}
where $a$ is fixed by the requirement that $\sigma_E/E = \num{5.8}\%$ at \SI{511}{keV}.
The comparison is shown in Fig.~\ref{FIG6B}, where the green points are from data and the blue line is from simulation.
The data points are scaled to the total rate observed before any cuts of \SI{26.6}{Hz}, which agrees well with the rate of \SI{25.9}{Hz} from simulation.
The contribution at S2 sizes larger than \SI{600}{keV} is still visible.

At energies below $\sim$\SI{150}{keV}, the simulation predicts a higher rate than observed in measurement.
This difference is due to a timing offset between the NaI(Tl) and the S1-signal, which causes a trigger on the falling edge of the S1 instead of on its peak amplitude and deteriorates the trigger efficiency for low energy recoils, as described in section~\ref{SEC2_3}.

\section{Single-electron S2-signals} \label{SEC4}
A distinct signal that is found in dual-phase xenon TPCs is that of S2-signals produced by single electrons \cite{Edwards200854,Santos2011,0954-3899-41-3-035201}.
The scintillation light of xenon, at \SI{178}{\nano m}, can liberate electrons in the TPC.
In general, the electrons come from impurities in the xenon that have a low ionization potential, such as O$^{-}$~ions, or from exposed metallic surfaces. 
If these electrons are somewhere in the active volume, they will in turn drift upward and create very small S2-signals.
Since the main S2-signal is the dominant source of UV photons in the TPC, it causes the large majority of single-electron peaks.

\begin{figure}[h]
\begin{center}
\includegraphics[width= \linewidth]{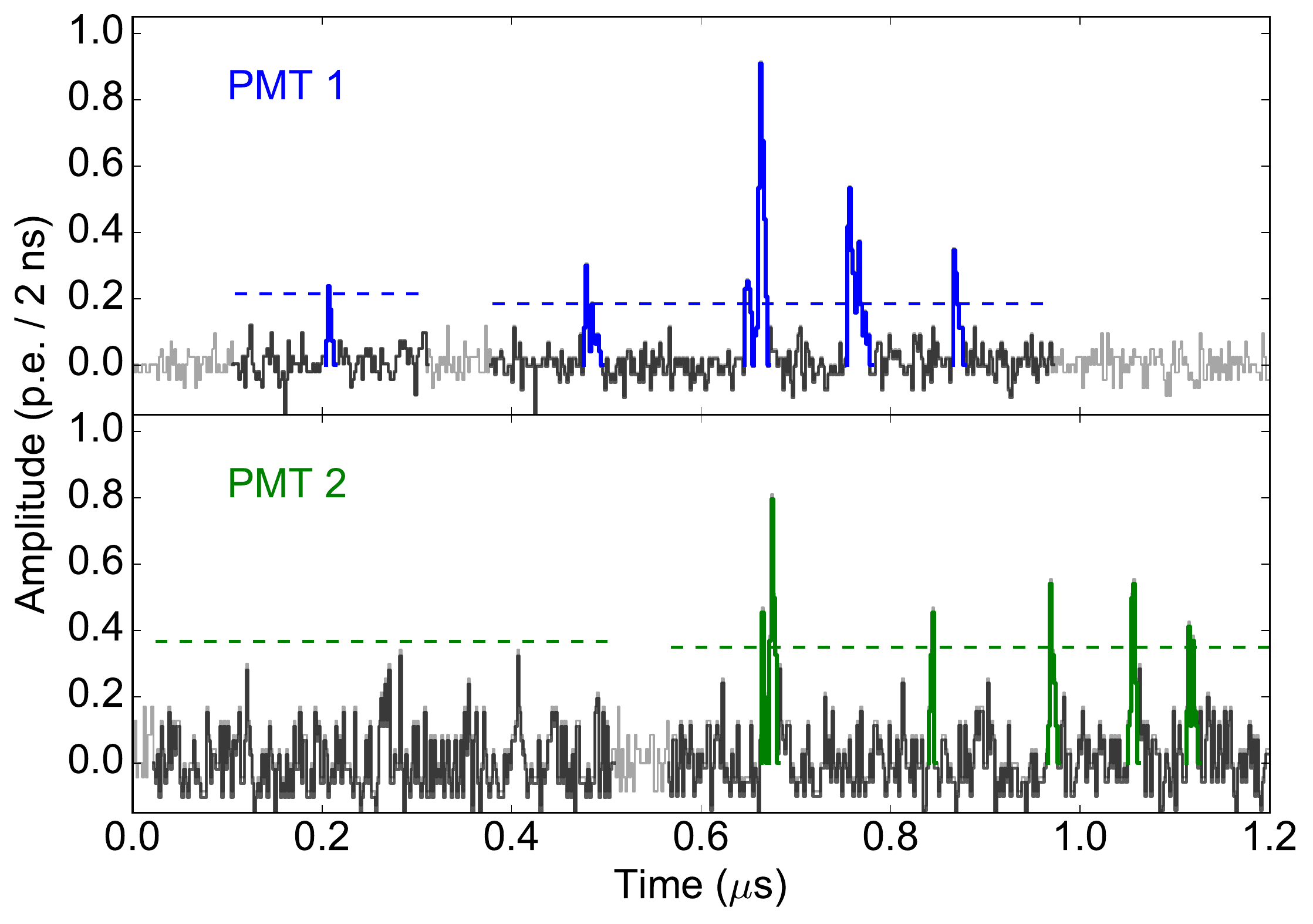}
\caption{
Example of a single-electron S2-signal, shown for both PMTs separately.
The data is cut into small pieces based on the crossing of a very low threshold.
This is indicated by the dark gray part of the waveforms.
The noise level is determined on the first 40~samples of these pieces, yielding a hitfinder threshold of \num{4.5} times the standard deviation of the noise (indicated by the dashed lines).
When the waveform crossed this threshold, a hit is found, indicated by the blue and green waveforms.
}
\label{FIG7}
\end{center}
\end{figure}

Fig.~\ref{FIG7} shows an example of a single-electron signal found in the data.
As described in section~\ref{SEC3_1}, the waveform is cut into small sections analogous to zero-length encoded data, so that the hitfinder threshold is dynamically updated based on the local noise level.
The blue and green parts of the waveform show the hits that are found, when a threshold of \num{4.5} times the standard deviation of the baseline noise is crossed (indicated by the dashed lines).
The width of the signal is around \SI{1}{\micro s}, comparable to normal S2-signals.

\begin{figure}[h]
\begin{center}
\includegraphics[width= \linewidth]{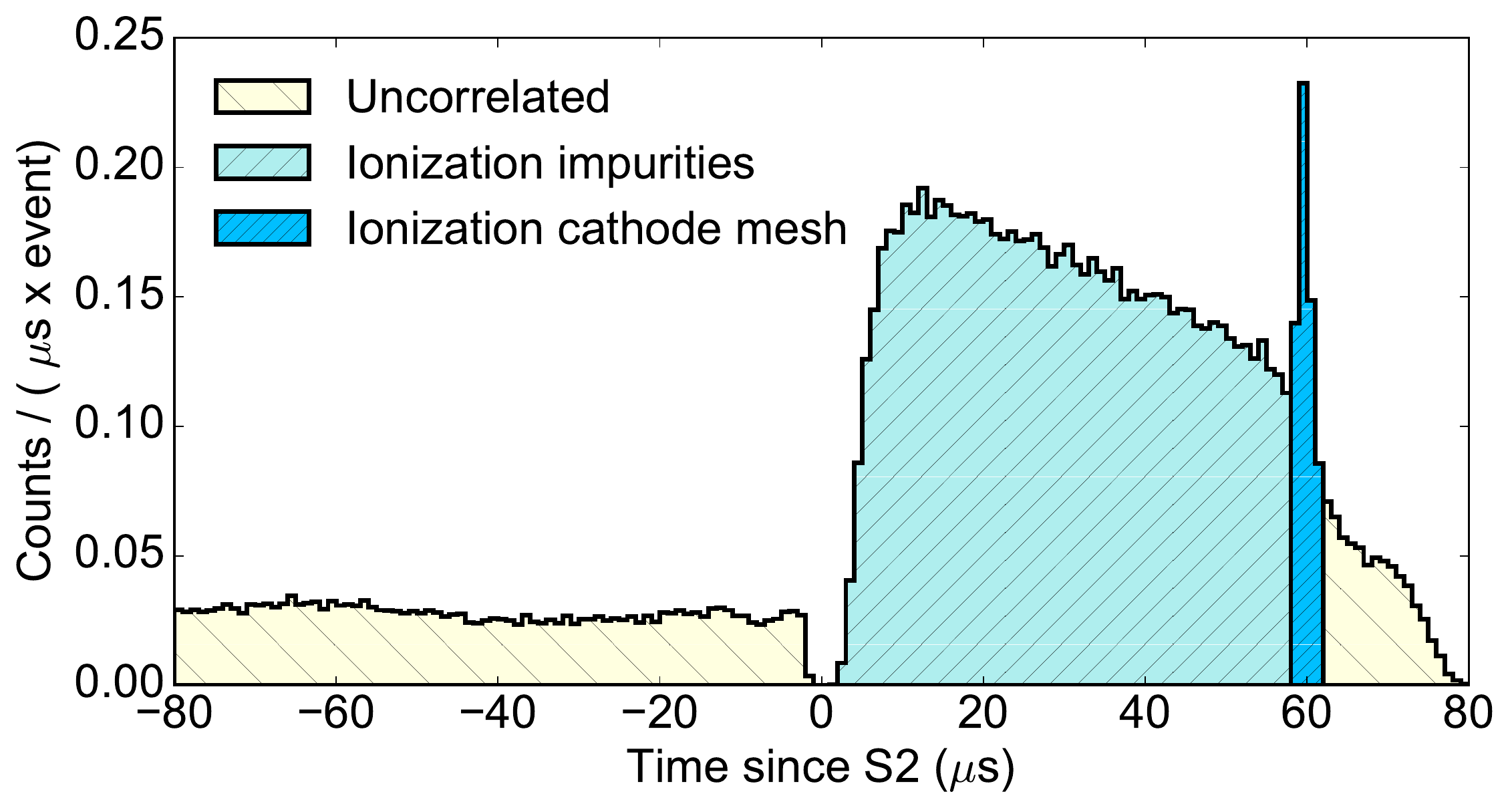}
\caption{Distribution of peak positions relative to the position of the S2. 
A large fraction of the peaks occurs between \SIrange{0}{61}{\micro s}, which is expected for single-electron signals that are caused by the S2 light.
The large peak at \SI{60}{\micro s} is due to the cathode mesh at this drift length: electrons are easily liberated from the stainless steel.}
\label{FIG8}
\end{center}
\end{figure}

Fig.~\ref{FIG8} shows the time distribution of the peak position relative to the position of the S2 for all candidate single-electron S2-signals, namely all peaks that are not classified as S1 or S2 and have a coincidence of both PMTs.
A large fraction of the peaks occurs between \SIrange{0}{60}{\micro s} (as defined by the maximum drift time).
We observe a clear increase at \SI{60}{\micro s}, which is due to the S2 light impinging on the cathode mesh, where electrons are liberated relatively easily due to the low ionization potential of the iron in the stainless steel.
Before the S2 ($\Delta t <$~\num{0}), as well as after the full drift length ($\Delta t >$~\SI{60}{\micro s}), there is a nonzero contribution, which is partly due to noise hits clustered into peaks, but partly shows the same properties as the single-electron signals in the drift region. 
These peaks can be caused by a delayed extraction phenomenon, as discussed in \cite{Santos2011} and \cite{0954-3899-41-3-035201}.

Single-electron S2s can be effectively used as `calibration sources': the detector response to just one electron can be probed in this way.
This enables the direct determination of various parameters, such as the secondary scintillation gain.
In addition, these signals can be used for a PMT gain calibration, since they consist of single-photoelectron hits.

\subsection{Gain calibration}
The PMT gain is defined as the average number of electrons at the anode responding to one electron emerging from the photocathode.
Often PMT calibrations are done using external pulsed light sources.
Although such calibration provides a direct and usually accurate gain calibration, it requires an interruption during dark matter data taking. 
In addition, a dedicated LED calibration system and calibration measurements are necessary.
Finally, the LED calibration is usually performed at a higher wavelength than the xenon scintillation light of \SI{178}{nm}, because it is technically challenging to guide UV light through an optical fiber.
This makes it impossible to probe effects like double-photoelectron emission, which occurs only at short wavelengths \cite{1748-0221-10-09-P09010}.

In this section, we discuss a method to use single-electron peaks for PMT gain calibration. 
These consist of well-separated single-photon hits and are abundant in all data, so they can be used to measure the PMT gain continuously. 
The LUX collaboration already uses single-electron signals as part of their gain calibration, which operates on different principles~\cite{new_lux}.

\subsubsection{Hit data selection}
Single-electron S2s typically have the same width as ordinary S2-signals (about \SI{1}{\micro s} wide), but consist of a small number of photoelectrons.
For example, for XENON100, these signals consists of roughly 20 photoelectrons \cite{0954-3899-41-3-035201}.

Since the hits in single-electron S2s are spread out over a relatively long duration, the PMT hits of the detected photons can be found individually (see Fig.~\ref{FIG7}).
This means that the single-electron S2s provide a source of single photoelectron hits, which can be used for an \emph{in-situ} gain calibration.

We apply cuts on the event, peak and hit level to select proper single-photon hits in proper single-electron signals.
Events are selected by the same criteria as in section~\ref{SEC3}.
For the peaks, defined as clusters of hits, we introduce the following cuts.
Since single-electron S2s are primarily caused by S2 photons, only peaks that occur within \SIrange{5}{60}{\micro s} after the S2 are selected. 
The lower bound ensures no fragments from accidentally split S2s are selected, and the upper bound cuts peaks beyond the maximum drift time.
Both PMTs are required to contribute to the peak, to suppress peaks consisting of noise and dark current hits.
In order to reduce contamination from common-mode noise clustering, we apply a cut on the average area of the hits in a peak: since noise hits on average have a smaller area than particle-induced PMT hits, a cluster of noise hits will have a low average area.

Finally, at the hit level, we do not consider hits with an extremely small width, indicative of noise hits rather than real PMT hits.
The width parameter used here is the sum absolute deviation~(SAD), given by
\begin{equation}
\rm{SAD} = \sum_i \frac{A_i}{A_{tot}} \left| t_i-t_c \right| {\rm ,}
\end{equation}
where $i$ runs over all samples in the hit, $t_c$ is the amplitude-weighted mean time of the hit, and $A$ denotes the area.
This parameter takes continuous values greater than or equal to zero, which has the advantage of discriminating different widths even if this is at the same order as the sampling time.
Typical values for a single-photoelectron hit are about \SI{3.5}{ns} for the XAMS PMTs.
We cut hits with an SAD less than \SI{0.5}{ns}, which mostly consist of hits that are just one sample wide (such that $\rm{SAD} = 0$ exactly) or where the hit is two samples wide but the area is dominated by just one sample.

With the above selection of hits we proceed to calibrate the gain of each PMT.
For gain calibration, the parameter of interest is the \emph{area} of the hits (given by $\int V dt$), which is proportional to the number of electrons~$n_e$ at the PMT output according to 
\begin{equation}
n_e = Q/q_e = \frac{1}{q_e} \int I dt = \frac{1}{q_e R} \int V dt {\rm ,}
\end{equation}
where $Q$ is the charge at anode, $q_e$ is the charge of the electron and where $R$ denotes the termination resistance of the digitizer (\SI{50}{\ohm}).
Because of this relation, the area of a hit can be expressed as number of electrons equivalent area.
If the hit results from one photoelectron, the average number of electrons at the output is simply $\mean{n_e} = G$, and the gain~$G$ can be computed.

\subsubsection{Acceptance correction} \label{SEC4_1_2}
\begin{figure}
\begin{center}
\includegraphics[width = \linewidth ]{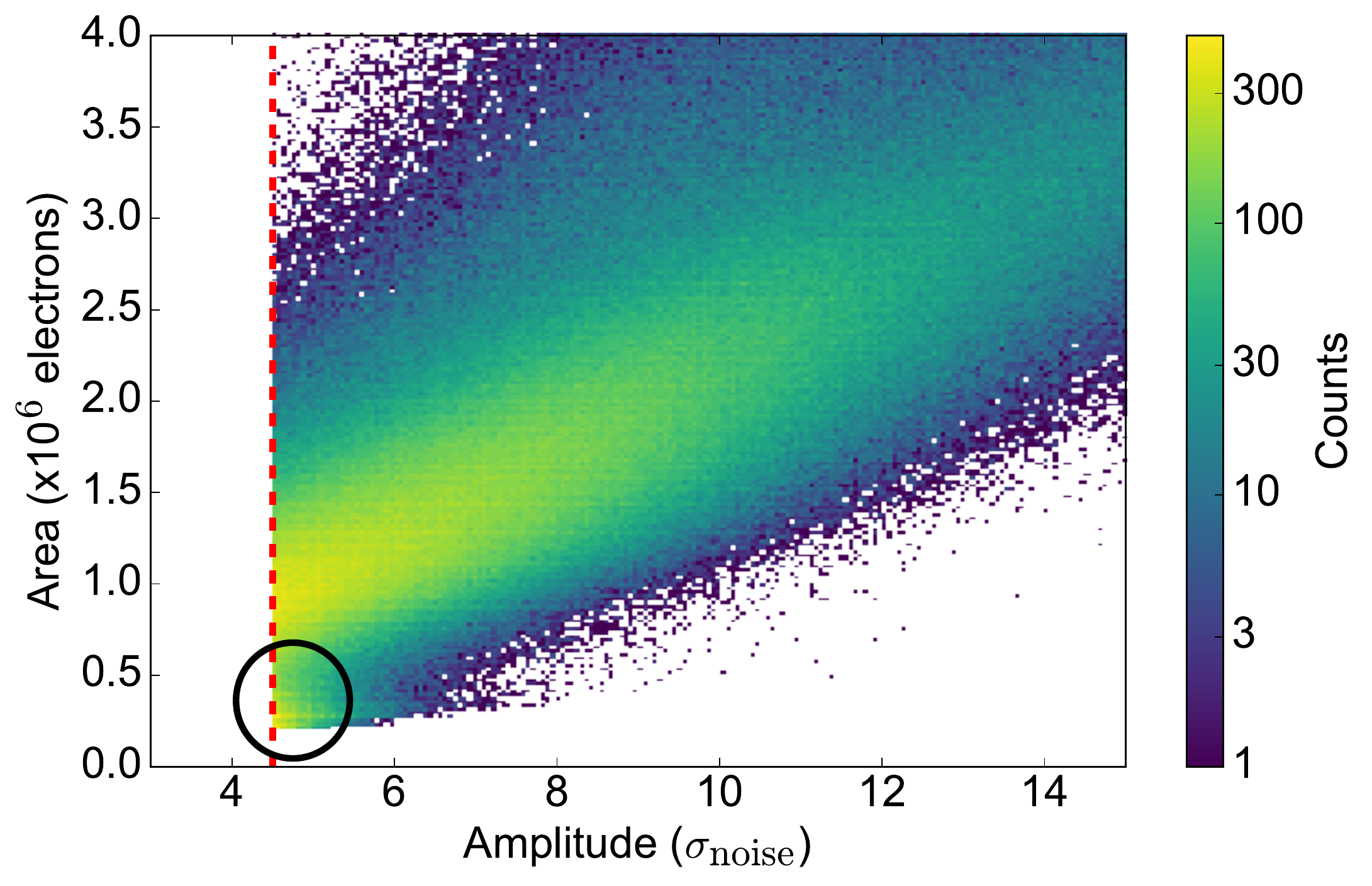}
\caption{The distribution of selected single-electron hits in amplitude and area. 
There is a clear correlation between the hit area and amplitude.
The distribution is cut at \num{4.5}$\sigma_{\rm noise}$ as defined by the hitfinding threshold.
A correction for hits below this threshold is calculated and used in the analysis.
At small area and low amplitude, a second band coming from noise hits can be seen (indicated by the black circle).
}
\label{FIG9}
\end{center}
\end{figure}

The hitfinding algorithm preferentially detects high-area hits, since these are more likely to exceed the hitfinding amplitude threshold.
Fig.~\ref{FIG9} shows the correlation between the amplitude, measured in units of $\sigma_{\rm noise}$, and the area of the hit.
The distribution is sharply cut at \num{4.5}$\sigma_{\rm noise}$, the hitfinder threshold.
This was chosen to limit the contribution of noise hits, which are visible in the bottom left corner of Fig.~\ref{FIG9}.

\begin{figure}[h]
\begin{center}
\includegraphics[width= \linewidth]{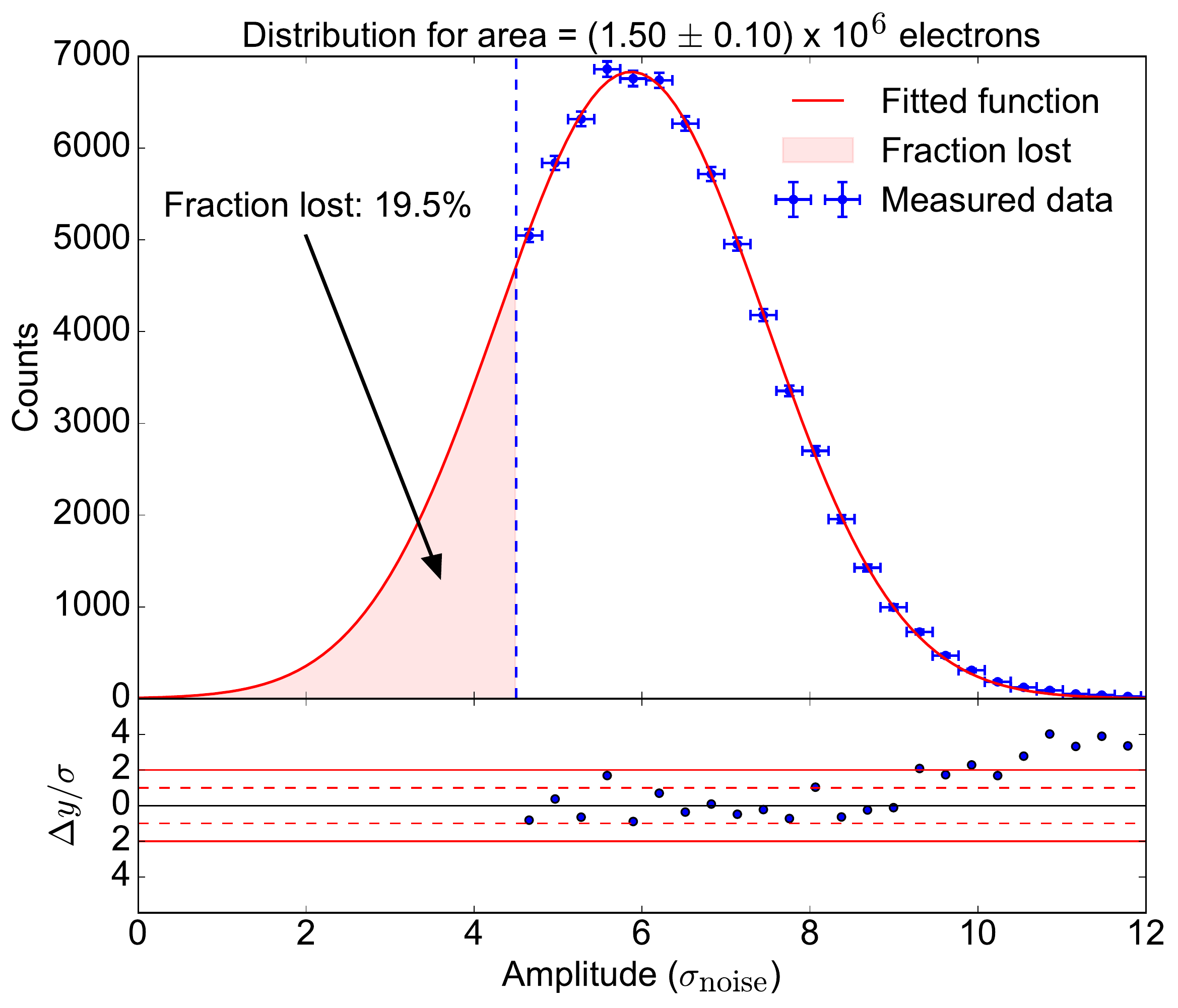}
\caption{Example of a fit used to determine the acceptance for an area slice. 
The fraction of the hits that is not found is inferred from the area under the fit in the region of the function that extends below the threshold. In this example, this fraction is \num{19.5}\%, so the acceptance is \num{80.5}\%.
The bottom panel shows the deviation from this fit in units of the error on the points.
The dashed and solid lines indicate a deviation of $1\sigma$ and $2 \sigma$, respectively.
Up to a threshold of \num{9}$\sigma_{\rm noise}$, the distribution is well described by the fit.
}
\label{FIG10}
\end{center}
\end{figure}

To correct for this loss of hits below the threshold, we must estimate the acceptance $\epsilon$ of the hitfinder, i.e.\ the fraction of photon hits found by the hitfinder, as a function of the hit area.
This function can be estimated by studying the amplitude distribution for a sample of hits with  similar area (equal up to a difference of \num{0.1e6}~electrons).
This distribution is cut at the hitfinder threshold, but since it is well-described by a Gaussian distribution above the threshold, we will assume that it follows a Gaussian function also below this threshold.
By evaluating the fraction of the area under the fitted distribution below the threshold, the acceptance can be calculated, as illustrated in Fig.~\ref{FIG10}.

\begin{figure}[h]
\begin{center}
\includegraphics[width= \linewidth]{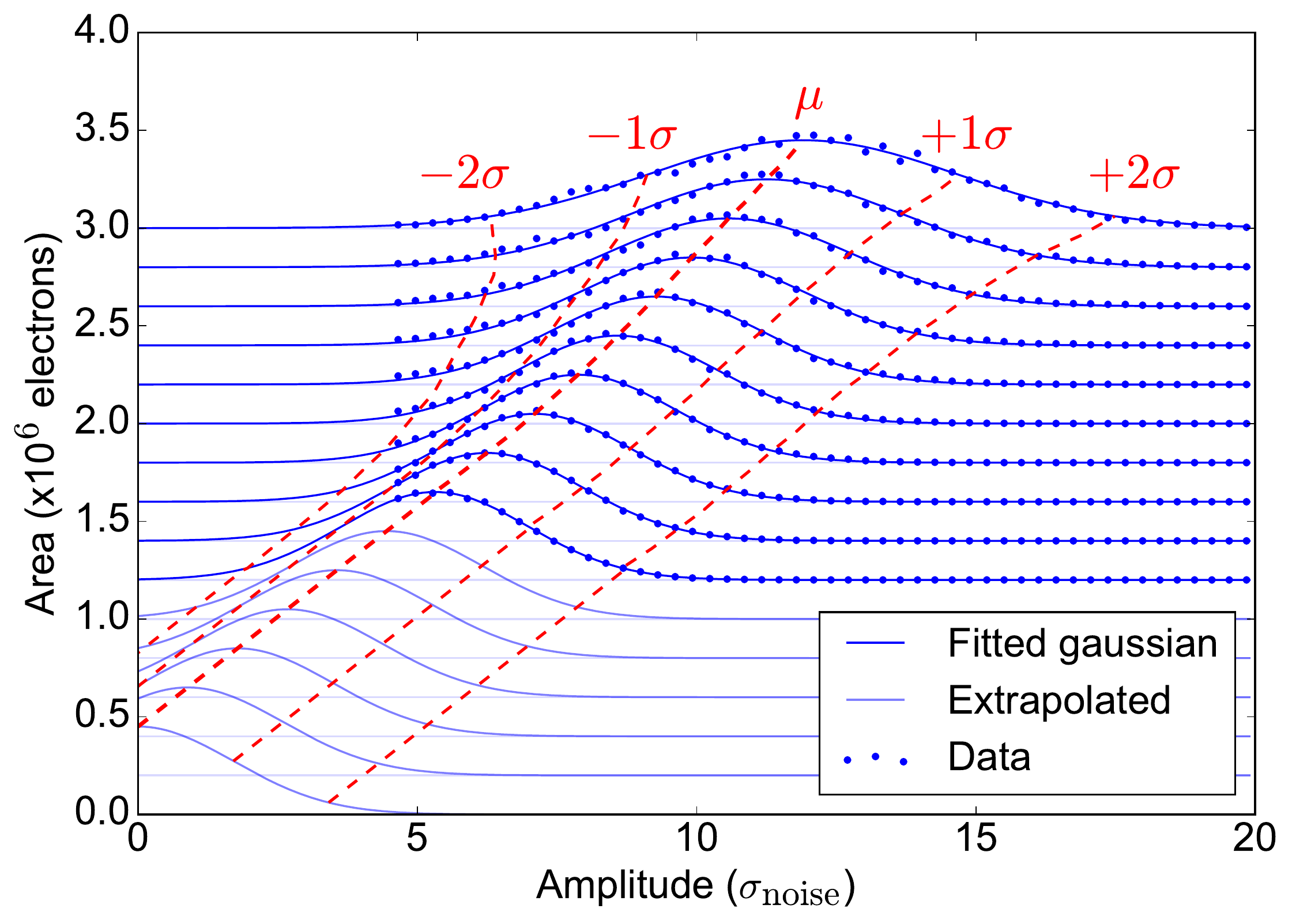}
\caption{
Stacked hit amplitude histograms for each area slice (blue points), together with Gaussian fits (blue lines). 
The data as well as the fits are scaled such that the maximum amplitude of all distributions is the same.
Red dashed lines indicate the mean and standard deviations of the fitted Gaussians. 
For low area slices, the amplitude distribution is estimated by extrapolating the mean and standard deviations found in higher-area slices as described in the text.}
\label{FIG11}
\end{center}
\end{figure}

The distribution of hits and Gaussian fits are shown for all area slices in Fig.~\ref{FIG11}.
For low-area hits, only a tail of the Gaussian distribution exceeds the hitfinding threshold of \num{4.5}$\sigma_{\rm noise}$, making it difficult to fit the distribution.
We instead infer the parameters $\mu$ and $\sigma$ by extrapolation.
Since the shape of PMT hits is to a good approximation independent of the area, the mean~$\mu$ is extended linearly to zero \cite{erik_master}.
We assume that the standard deviation~$\sigma$ is constant at low area, since this should be dominated by baseline noise on the highest bin and is therefore independent of the hit area.

With the amplitude distribution specified by $\mu(A)$ and $\sigma(A)$, the acceptance for every area and for different hitfinding thresholds can be computed.
For a hitfinding threshold of $n_{tr} \sigma$, the acceptance~$\epsilon$ as a function of peak area~$A$ is
\begin{equation}
\epsilon(A) = \int_{n_{tr} \sigma}^{\infty} g\left( x ; \mu(A), \sigma(A) \right) dx
\label{EQN5}
\end{equation}
where $g$ is a normalized Gaussian distribution and $x$ denotes the amplitude.
To correct the area spectrum, it is divided by~$\epsilon(A)$.

\begin{figure}[h]
\begin{center}
\includegraphics[width=\linewidth]{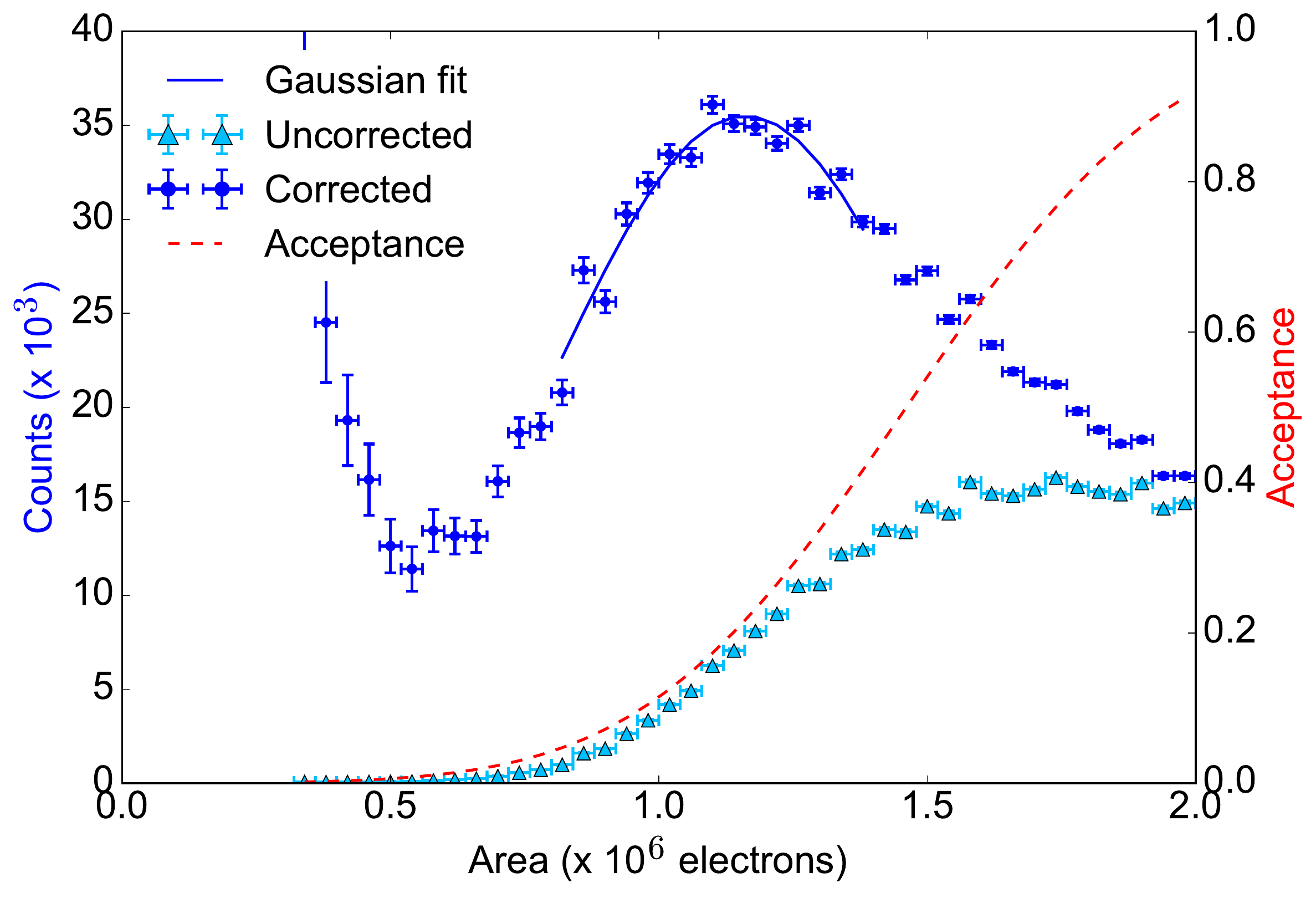}
\caption{
The area distribution for the hits with (dark blue points, left scale) and without (light blue points, left scale) correction from the acceptance function (red dashed line, right scale) as determined from the model described in section~\ref{SEC4_1_2}.
A Gaussian fit (blue solid line) is used to determine the gain.
At low area, noise hits give a large contribution to the corrected spectrum, since they are highly amplified by the acceptance correction.
Points below \num{0.3e6} electrons area, where the acceptance drops below \num{0.1}\%, are omitted from the plot.
}
\label{FIG12}
\end{center}
\end{figure}

Although a hitfinder threshold that is as low as possible is desired for determining the acceptance, it is not necessarily ideal for determining the gain.
This is because low-amplitude noise hits are too dominant in the area spectrum, so that the gain will be underestimated.
We therefore use a higher amplitude threshold to remove any possible bias due to noise hits, which we compensate by a corresponding change in the acceptance function.
In Fig.~\ref{FIG12}, the uncorrected and corrected area spectrum for a hitfinding threshold of 6.5$\sigma_{\rm noise}$ is shown, together with the acceptance function for this threshold.
A clear peak is visible, which is fit in the area around the peak to determine the gain.
The fit is limited to a part around the maximum; at low area, the noise contribution becomes dominant, while at high area the contribution from two-photoelectron hits cannot be excluded.
Similar features are found in other PMT calibrations, such as in~\cite{1502.01000}.

\begin{figure}[h]
\begin{center}
\includegraphics[width= \linewidth]{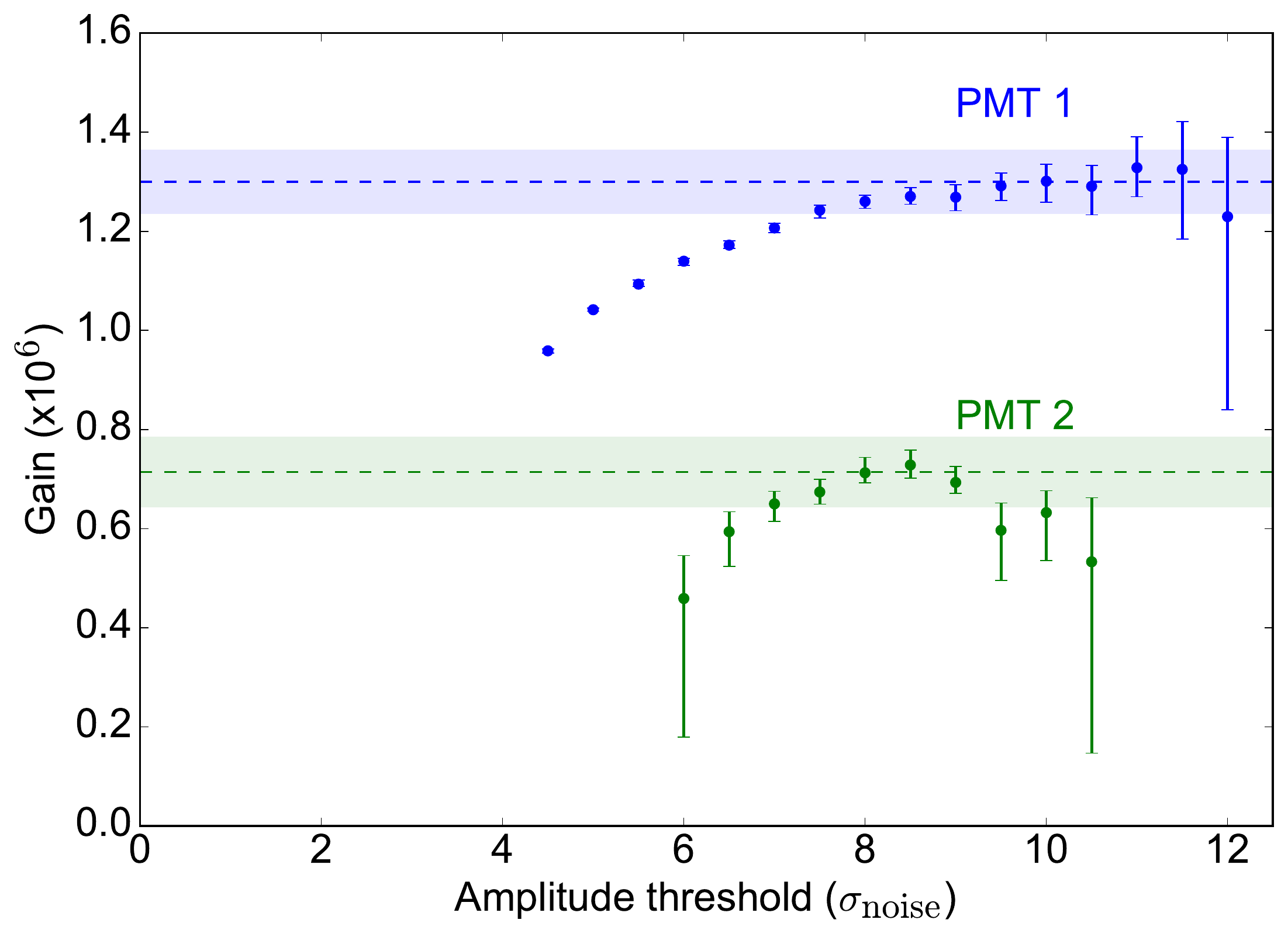}
\caption{The gain for different hitfinding thresholds (in units of standard deviation of the noise), for the two PMTs in XAMS. 
As the hitfinding thresholds rises far enough above the low-amplitude noise, the gain that is determined levels off to one value, which is the gain of the PMT.
The error bars are calculated by allowing varying values for $\mu(A)$ and $\sigma(A)$ for the acceptance function.
At high thresholds, the result of the fit is very sensitive to variations in the acceptance function, causing the large error bars.
The systematical errors are estimated to be 5\% for PMT~1 and 10\% for PMT~2, as shown with the bands, based on the convergence of the final points.
}
\label{FIG13}
\end{center}
\end{figure}

In Fig.~\ref{FIG13}, the determined gain is plotted as a function of the hitfinding threshold.
For low thresholds, the noise contribution becomes too pronounced to properly fit the spectrum.
For higher thresholds, the gain that is determined converges to a final value, which we infer to be the true gain of the PMT.
We allow a range around this value from uncertainty on the convergence of the final points, which we estimate to be 5\% for PMT~1 and 10\% for PMT~2.
The PMT gains found in this analysis were $(1.30 \pm 0.07) \times 10^6$ for PMT~1 and $(0.71 \pm 0.07) \times 10^6$ for PMT~2; both close to the typical gain of \num{1.0e6} quoted by Hamamatsu for this type of PMT \cite{hamamatsu}.
The error bars in Fig.~\ref{FIG13} originate from systematic errors on the acceptance function, which we calculate by perturbing the fit parameters $\mu(A)$ and $\sigma(A)$ in equation~\ref{EQN5}.

\subsection{Discussion}
The method used in this work relies on modeling the PMT hits.
In particular, the amplitude of hits of a given area is assumed to be normally distributed.
For large-area hits, this assumption can be verified since most hits are above the hitfinding threshold, and the distribution follows a Gaussian distribution to a high degree.
For smaller areas, a significant part of the distribution is inaccessible and needs to be inferred from the visible part of the distribution (as in Fig.~\ref{FIG10}).
Moreover, the parameters~$\mu$ and~$\sigma$ of the distribution are extrapolated from larger areas where fits can be made (Fig.~\ref{FIG11}).
The validity of the extrapolation can break down at small area, although this will not affect the gain determination if the approximation is valid sufficiently far below the average single-photoelectron hit area.

In the case of low PMT gains, the separation of noise and true PMT hits becomes a serious issue.
This is the case for PMT~2 in our analysis, where at high thresholds the errors increase and we eventually fail to fit the distribution because of low statistics.
Moreover, we cannot confirm if convergence is reached before this effect starts to dominate.
We therefore estimate the systematic errors to be higher for PMT~2 than for PMT~1.
It should be noted that these limitations becomes less important if the PMT gain is higher, so that the PMT hits are more separated from the noise.

A PMT calibration requires a source of single photoelectron hits.
For single-electron S2s, a few photoelectron signals are seen in the PMT channels over a time window of typically \SI{1}{\micro s}.
There is a finite probability of having two or more PMT signals clustered together into one hit, so that there is a contribution of two-photoelectron hits in the data.
The importance of this effect could be different for other TPCs, as it depends on several parameters such as the transient time spread of the PMTs, the sampling time of the ADCs, the width of the S2 and the anode voltage.
These effects will thus need to be studied further if this method is to be used for other TPCs.

Compared to the normal PMT calibration with LED pulsed light, there are some definite advantages to using single-electron S2s.
One of these is they are usually readily available and easily identified in ordinary (energy) calibration or dark matter data; no extra dedicated calibration runs are required.
This means that the drift of PMT gains can be monitored on timescales far shorter than with ordinary PMT calibration runs.
A second advantage is that the response to the scintillation light is directly probed.
The scintillation light of xenon has a wavelength of \SI{178}{nm}, but since this is technically challenging to provide for a calibration, higher wavelengths are used.
For example, XENON100 uses an LED at \SI{470}{nm} \cite{Akerib:2012ys}.
The method described here makes it possible to study, for example, the possibility of two-photoelectron emission due to one scintillation photon at the photocathode.

\section{Summary} \label{SEC5}
In this work, the first data of the XAMS TPC were presented.
An energy resolution of (\num[separate-uncertainty = true]{5.8 \pm 0.2})\% was achieved at \SI{511}{keV}.
The electron lifetime was found to be \SI[separate-uncertainty = true]{429 \pm 26}{\micro s}, which is sufficient for this TPC, after only 7 days of purification.
An average light yield of (\num[separate-uncertainty = true]{5.6 \pm 0.3})~photoelectrons/\si{keV} (recalculated to zero field and \SI{122}{keV}) was found, which is comparable to TPCs like XENON100 and LUX.

A new PMT calibration method based on single-electron S2-signals was explored.
Since single-electron S2-signals are very abundant in dual-phase xenon TPCs, this method of PMT calibration can give an important independent cross-check of the normal PMT calibration, with the advantage of superior time resolution and no need for dedicated PMT calibration data.

\section{Acknowledgments}
This work is part of the research program of the Foundation for Fundamental Research on Matter (FOM), which is part of the Netherlands Organization for Scientific Research (NWO).
We gratefully acknowledge the technical support from to the mechanical, electrical and computing departments at Nikhef.

\section{References}

\bibliography{references}

\appendix
\newpage

\onecolumn
\section{Piping and instrumentation diagram} \label{APP_A}

\begin{figure}[!h]
\begin{center}
\includegraphics[width= \linewidth]{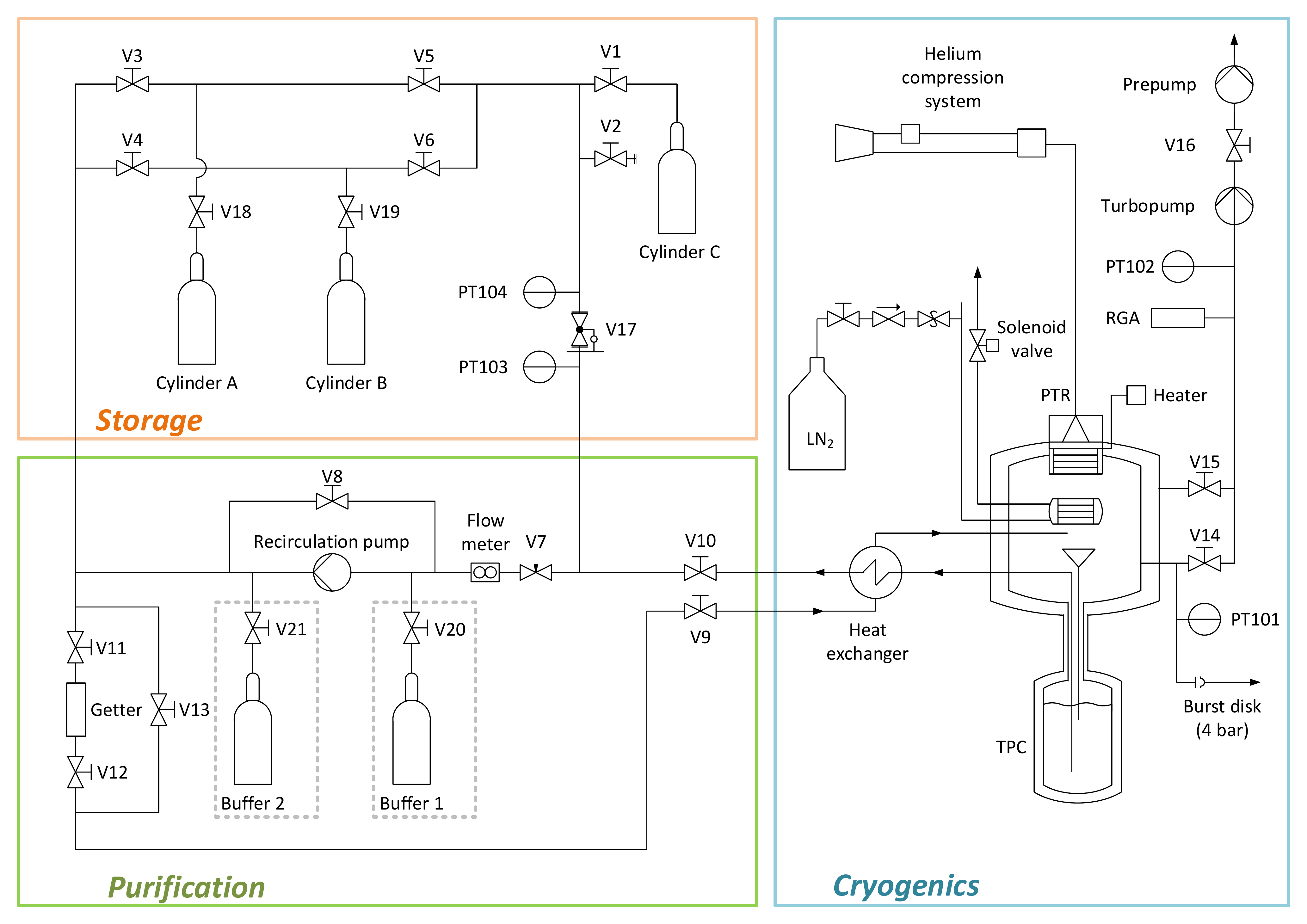}
\caption{
Piping and instrumentation diagram of the setup. 
In the box on the right, the cryogenic part of the setup is shown.
The main cooling is provided by the pulse tube refrigerator~(PTR).
The liquefied xenon drops down into a funnel, which leads into the detector volume.
The temperature at the cold finger is controlled by adjusting the current to a resistive heating band.
In case of a cooling failure, automatic emergency cooling is provided by a pressurized liquid nitrogen system.
The bottom left box shows the purification system.
Liquid xenon is extracted from the TPC and evaporates in the heat exchanger, so that gaseous xenon is be pumped through a getter.
The flow speed is regulated with a needle valve (V7) and measured with a mass flow meter.
The buffer volumes, shown in the dashed gray boxes, were added in the system after the measurements.
When the detector is not running, we store the xenon as a pressurized gas in the storage system, shown in the top left.
A pressure regulator (V17) serves to set a low pressure in the detector volume.
Cylinders~A and~B can be submerged in liquid nitrogen dewars, causing xenon deposition on in the cylinders.
This is used to recuperate the xenon from the setup.
}
\label{FIG14}
\end{center}
\end{figure}

\end{document}